\documentclass[aps,reprint,nofootinbib,preprintnumbers,twocolumn,floatfix,superscriptaddress,prapplied,showpacs,longbibliography]{revtex4-1}
\usepackage{physics}
\usepackage{natbib}
\usepackage{amsmath,amssymb}
\usepackage{graphicx}
\usepackage[dvipsnames]{xcolor}
\usepackage[caption=false]{subfig}
\usepackage{float}
\usepackage[pdftex]{hyperref} %last command in preamble
\usepackage{qcircuit}
\usepackage{wasysym}
\usepackage{dsfont}
\usepackage{bm}

\usepackage{dsfont}
\newcommand{\Id}{\ensuremath{\mathds{1}}} % for the Identity operator
\newcommand{\rot}[1]{\ensuremath{\textsf{R}_{#1}}}
\newcommand{\Lspam}{\ensuremath{\mathcal{L}_\text{SPAM}}}

\usepackage{array}
\newcolumntype{P}[1]{>{\centering\arraybackslash}p{#1}} % formatting for Table 1

\usepackage[normalem]{ulem}
\usepackage{cancel}

\usepackage[ruled,vlined]{algorithm2e}

\begin{document}
	\title{Lindblad Tomography of a Superconducting Quantum Processor}
	
	\author{Gabriel O. Samach}
	\email[Corresponding author: ]{ga26317@mit.edu}
	\affiliation{Research Laboratory of Electronics, Massachusetts Institute of Technology, Cambridge, MA 02139}
	\affiliation{MIT Lincoln Laboratory, Lexington, MA 02421}

	\author{Ami Greene}
	\affiliation{Research Laboratory of Electronics, Massachusetts Institute of Technology, Cambridge, MA 02139}

	\author{Johannes Borregaard}
	\email[Corresponding author: ]{j.borregaard@tudelft.nl}
	\affiliation{Qutech and Kavli Institute of Nanoscience, Delft University of Technology, Delft, The Netherlands}
	\affiliation{Department of Mathematical Sciences, University of Copenhagen, Copenhagen, Denmark}

	\author{Matthias Christandl}
	\affiliation{Department of Mathematical Sciences, University of Copenhagen, Copenhagen, Denmark}

	\author{\\Joseph Barreto}
	\affiliation{Qutech and Kavli Institute of Nanoscience, Delft University of Technology, Delft, The Netherlands}

	\author{David K. Kim}
	\affiliation{MIT Lincoln Laboratory, Lexington, MA 02421}

	\author{Christopher M. McNally}
	\affiliation{Research Laboratory of Electronics, Massachusetts Institute of Technology, Cambridge, MA 02139}

	\author{Alexander Melville}
	\affiliation{MIT Lincoln Laboratory, Lexington, MA 02421}

	\author{Bethany M. Niedzielski}
	\affiliation{MIT Lincoln Laboratory, Lexington, MA 02421}

	\author{\\Youngkyu Sung}
	\affiliation{Research Laboratory of Electronics, Massachusetts Institute of Technology, Cambridge, MA 02139}

	\author{Danna Rosenberg}
	\affiliation{MIT Lincoln Laboratory, Lexington, MA 02421}

	\author{Mollie E. Schwartz}
	\affiliation{MIT Lincoln Laboratory, Lexington, MA 02421}

	\author{Jonilyn L. Yoder}
	\affiliation{MIT Lincoln Laboratory, Lexington, MA 02421}

	\author{Terry P. Orlando}
	\affiliation{Research Laboratory of Electronics, Massachusetts Institute of Technology, Cambridge, MA 02139}

	\author{Joel I-Jan Wang}
	\affiliation{Research Laboratory of Electronics, Massachusetts Institute of Technology, Cambridge, MA 02139}

	\author{Simon Gustavsson}
	\affiliation{Research Laboratory of Electronics, Massachusetts Institute of Technology, Cambridge, MA 02139}

	\author{Morten Kjaergaard}
	\affiliation{Research Laboratory of Electronics, Massachusetts Institute of Technology, Cambridge, MA 02139}
	\affiliation{Center for Quantum Devices, Niels Bohr Institute, University of Copenhagen, Copenhagen, Denmark}
	
	\author{William D. Oliver}
	\affiliation{Research Laboratory of Electronics, Massachusetts Institute of Technology, Cambridge, MA 02139}
	\affiliation{MIT Lincoln Laboratory, Lexington, MA 02421}
	\affiliation{Department of Electrical Engineering and Computer Science, Massachusetts Institute of Technology, Cambridge, MA 02139}

	\date{\today}

	\begin{abstract}
    As progress is made towards the first generation of error-corrected quantum computers, robust characterization and validation protocols are required to assess the noise environments of physical quantum processors.  While standard coherence metrics and characterization protocols such as $T_{1}$ and $T_{2}$, process tomography, and randomized benchmarking are now ubiquitous, these techniques provide only partial information about the dynamic multi-qubit loss channels responsible for processor errors, which can be described more fully by a Lindblad operator in the master equation formalism.  Here, we introduce and experimentally demonstrate Lindblad tomography, a hardware-agnostic characterization protocol for tomographically reconstructing the Hamiltonian and Lindblad operators of a quantum noise environment from an ensemble of time-domain measurements.  Performing Lindblad tomography on a small superconducting quantum processor, we show that this technique naturally builds on standard process tomography and $T_{1}$/$T_{2}$ measurement protocols, characterizes and accounts for state-preparation and measurement (SPAM) errors, and allows one to place bounds on the fit to a Markovian model  Comparing the results of single- and two-qubit measurements on a superconducting quantum processor, we demonstrate that Lindblad tomography can also be used to identify and quantify sources of crosstalk on quantum processors, such as the presence of always-on qubit-qubit interactions.
	\end{abstract}
	\maketitle

% --------------------------------------------------------------------------------------
\section{Introduction}
Quantum computers offer computational power fundamentally distinct from that of their classical counterparts and are predicted to offer an advantage for certain problems in fields such as quantum chemistry and optimization, which are often intractable on even the largest classical supercomputers~\cite{Montanaro2016,Georgescu2014}. The promise of quantum advantage over classical hardware has driven extensive efforts to build quantum computing devices based on a number of different hardware platforms---including trapped ions~\cite{Figgatt2019,Nam2020}, neutral atoms~\cite{Keesling2019,Ebadi2021}, and superconducting circuits~\cite{Arute2019}---each of which is susceptible to characteristic imperfections and noise mechanisms which can limit performance. 

In order to mitigate these sources of error, fault-tolerant quantum error correction protocols encode logical qubits across many physical qubits, provided the error rate of the physical qubits is below a threshold~\cite{Aliferis2006,Aharonov2008}. This approach, however, comes with considerable overhead in terms of the additional qubits needed for the encoding~\cite{Iyer2018}. While the overhead required for generic device-agnostic error correction schemes may prove prohibitive in the near-term, the need for redundancy can be substantially reduced by tailoring the correction scheme to the specific noise environment and imperfections of the particular quantum processor~\cite{Tuckett2018,Stephens2013,Tuckett2019,Xu2019,Tuckett2020}.
% As progress is made towards the first generation of error-corrected quantum computers, we require robust measurement protocols for characterizing the performance and loss mechanisms of real quantum devices. 

To date, a broad toolbox of quantum characterization, verification, and validation (QCVV) techniques have been proposed and utilized which focus on different aspects of device performance---such as randomized benchmarking (RB)~\cite{Emerson2005,Magesan2011}, gate set tomography (GST)~\cite{blume_kohout2017,Nielsen2021,Proctor2020} and state/process tomography~\cite{Paris2004}---each with their own strengths and weaknesses~\cite{eisert_2020}. For example, randomized benchmarking provides an approach for assessing the average fidelity of quantum gate operations independent of state preparation and measurement (SPAM) errors, and it has consequently become a standard measure of performance for experimental quantum devices.  However, the average fidelity alone does not provide much information about the actual noise processes at play in the device, the details of which are crucial to more fully modeling the device and developing tailored error mitigation and correction techniques.

State and process tomography, on the other hand, provide more detailed information about discrete moments in a qubit's evolution, such as the qubit state at a particular time or the quantum process corresponding to a gate operation of a fixed duration.  However, caution must be exercised in order to consistently interpret the results of tomography in the presence of SPAM errors~\cite{Mogilevtsev2012}. Building on traditional tomographic protocols, a number of theoretical and experimental works have demonstrated self-consistent characterization of SPAM errors in process tomography and gate characterization~\cite{Mogilevtsev2012,Merkel2013,Medford2013,Kim2015}. Common to many of these techniques is the use of maximum likelihood estimation (MLE), which provides a robust and flexible estimation procedure capable of handling over-complete data and constrained problems. While such techniques offer a promising step forward, the characterization of a discrete moment in a qubit's evolution is not always sufficient, and one often requires detailed knowledge about how the noise environment and crosstalk between qubits dynamically influence evolution in time~\cite{Burnett2019}.

Here, we present a robust technique for characterizing the dynamics of a multi-qubit system from an ensemble of time-domain measurements, which we call Lindblad tomography (LT).  As a characterization tool, LT can be used to analyze any general time-independent and memory-less noisy quantum process.
% As a characterization tool, LT can be used to analyze any general time-independent quantum channel governing the evolution of a quantum system in time. 
For example, one could use LT to characterize the noise processes experienced by a qubit during free evolution---such as $T_1$- and $T_2$-processes, which can be formally described as amplitude damping and dephasing channels respectively. Similarly, LT could also be used to evaluate and diagnose a deliberately engineered channel, such as a tailored Hamiltonian implemented on an analogue quantum simulator~\cite{Hangleiter2021}. 
% Additionally, LT can be used to aid the characterization and calibration of quantum gates by performing chains of repeated gates which ideally cancel to unity (such as two-qubit \textsf{CZ} gates) and characterizing the deviation~\cite{Kjaergaard2020,Sung2021}.

The goal of LT is to estimate the Hamiltonian, quantum jump operators, and corresponding decay rates that describe the evolution of interest using maximum likelihood estimation (MLE), a process we collectively refer to as extracting the \emph{Lindbladian} of the channel. In doing so, we assume the channel can be well-approximated by a time-independent master equation. Prior to extracting the Lindbladian of the channel, our protocol uses a subset of measurement data to characterize the SPAM errors for the device, which we then include in our estimation of the Lindbladian from the full set of measurement data.  As such, we assume that the SPAM errors are constant across the full set of LT measurements, and thus are time-independent during the duration of data collection. To summarize, the main requirements for LT are thus:
\begin{enumerate}
\item The evolution of the quantum system should be Markovian and well described by a \emph{time-independent} master equation.
\item SPAM errors are assumed to be constant during the full duration of data acquisition.
\end{enumerate}
% In addition to these two fundamental assumptions, we note that further simplifications may be made based on specific hardware considerations---such as a third assumption of perfect single-qubit gates for LT of high fidelity superconducting qubits, discussed further below---though such assumptions are not required for performing LT in general.

\begin{figure}[t!]
\includegraphics{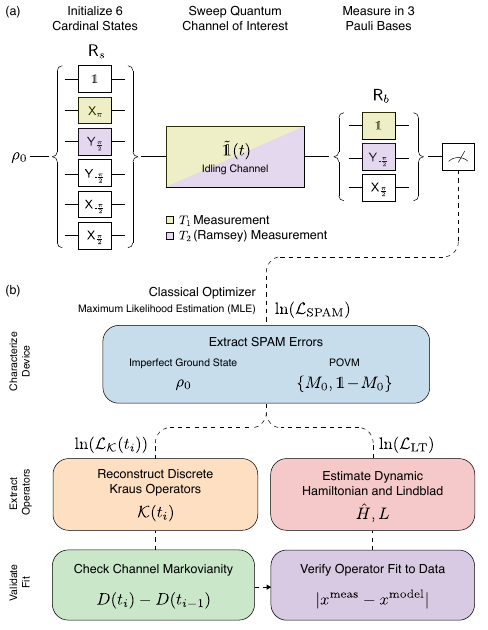}
\caption{Single-qubit Lindblad tomography (LT) protocol.  
	(a) The sequence of measurements required for single-qubit LT. The qubit is prepared in its imperfect ground state $\rho_0$ and one of six single-qubit pre-pulses $\rot{s}$ is applied to rotate the qubit as close as possible to each cardinal state of the Bloch sphere; free evolution of the quantum system is swept; and one of three post-pulses $\rot{b}$ is applied to rotate the measurement axis into each Pauli basis. 
	Notably, the set of pre- and post-pulses $\{\rot{s},\rot{b}\}$ includes all of the rotations required for standard process tomography, allowing one to reconstruct the channel at each discrete time step as in process tomography (b, left path), as well as for continuous time using all time steps (b, right path).
	(b) Analysis protocol for LT. Results from all combinations of pre-/post-pulses and channel durations are passed to a classical optimizer based on maximum likelihood estimation (MLE).  SPAM errors due to imperfect ground state preparation and measurement infidelity are extracted from data at $t=0$, and the results are used to separately estimate: (left path) the Kraus operators $\mathcal{K}(t_i)$ for each discrete channel duration $t_i$ and channel Markovianity using the trace distance $D$ between pairs of states; (right path) the Hamiltonian $\hat{H}$ and Lindblad matrix $L$ for continuous time $t$, where the operator fit to data is evaluated using the average error between the measurement outcomes predicted by the operators ($x^\text{model}$) and data ($x^\text{meas}$).
	}
	\label{fig:LT_structure}
\end{figure}

As we demonstrate, the set of measurements required for LT contains all the measurements required to perform process tomography at many time points, and this data can thus be used to independently extract the Kraus operators of the channel at discrete times. These operators can then be used to qualitatively validate the assumption of Markovianity using the measure proposed in Ref.~\cite{Breuer2009}. Performing LT on a superconducting quantum processor, we show that this verification technique enables us to identify potential sources of non-Markovianity which arise due to always-on crosstalk between neighboring qubits.

While estimation techniques for Lindblad noise operators have been proposed and demonstrated previously for a single-qubit solid-state~\cite{Howard2006} and trapped-ion system~\cite{Shapira2020}, the characterization reported in this work differs from these past demonstrations in its careful account of SPAM errors during the estimation, its assessment of the Markovianity of the channel, and its use of MLE. In this respect, Lindblad tomography has much in common with gate set tomography (GST) and can be viewed as a strategic, application-specific simplification of ``long-sequence GST''~\cite{Nielsen2021}. In long-sequence GST, the goal is to extract the SPAM-consistent process map for each of the physical operations in a quantum gate set from a long sequence of repeated operations; to impose physicality constraints, long-sequence GST then estimates the Lindbladian which generated each gate, and a discrete process is obtained by evaluating the dynamic operators for the fixed duration of the physical gate. Since the standard gate set typically includes a period of free evolution (which we refer to as the Idling gate below), a small subset of the sequences required for long-sequence GST are equivalent to the sequences required for LT: i.e., the sequences where the series of applied gates (the `germ'~\cite{blume_kohout2017}) consists entirely of repeated applications of the Idling operation. However, while GST requires many additional sequences in order to consistently characterize all the operations in the gate set relative to each other, LT substantially cuts down the number of required sequences by focusing only on a single process: the Idling operation. As such, while GST provides a more complete description of the full set of qubit operations, LT allows one to bypass much of the analytical complexity and experimental overhead of GST, at the cost of a more targeted characterization.

Like GST and most other tomographic protocols, such as standard process tomography, the number of measurements required for Lindblad tomography scales exponentially with the number of qubits. As such, we note that full characterization of a large quantum processor with Lindblad tomography remains experimentally impractical. However, as we discuss in the conclusion, careful characterization of single- and two-qubit plaquettes across a device may prove sufficient to diagnose sources of qubit-qubit crosstalk and bootstrap higher order multi-qubit errors~\cite{Govia2020,Lilly2020}.
% However, as we discuss in the conclusion, careful characterization of two-qubit plaquettes across a device may prove sufficient for diagnosing sources of qubit-qubit crosstalk---a dominant source of non-Markovian errors in experimental quantum processors, as we show---the details of which can generically decomposed into two-body interactions between pairs of qubits.

The paper is organized as follows. In Section~\ref{sec:protocol}, we introduce the general technical framework behind Lindblad tomography. Applying LT to a small superconducting quantum processor, we then continue with a characterization of SPAM errors for this device in Section~\ref{subsec:spam}, estimation of Kraus operators and the degree of Markovianity in Sections~\ref{subsec:process_map} and~\ref{sec:markvovianity}, and finally Hamiltonian and Lindblad estimation in Section~\ref{subsec:lindblad_extraction}. In each section, we first present the hardware-agnostic protocol, and then consider the results of running the protocol on a device of coupled superconducting qubits.

% --------------------------------------------------------------------------------------
\section{Single-Qubit Lindblad Tomography Measurement Protocol} \label{sec:protocol}
 We first introduce Lindblad tomography in the context of characterizing a single qubit. The generalization to two or more qubits follows readily, as discussed below.   

The structure of single-qubit LT is illustrated in Fig.~\ref{fig:LT_structure}. To determine the Lindbladian describing a single-qubit channel, we perform the following over-complete set of single-qubit rotations and basis measurements (Fig.~\ref{fig:LT_structure}a):

\begin{enumerate}
    \item The qubit is initialized in a state $\rho_0$ close to its ground state and one of six single-qubit gates $\rot{s} = \{\Id, \textsf{X}_{\pi}, \textsf{Y}_{\pm\!\frac{\pi}{2}}, \textsf{X}_{\mp\!\frac{\pi}{2}} \}$ is applied, initializing the qubit as close as possible to each of the six cardinal states of the Bloch sphere ($\ket{0}$, $\ket{1}$, $\ket{\pm}$, $\ket{\pm i}$) respectively.
    \item The idling channel $\tilde{\Id}(t)$ is swept as a function of time, corresponding to a variable time delay between state preparation and measurement during which no experimental controls are performed on the qubit. In the absence of any noise, the idling channel would correspond to the identity channel $\Id(t)$. 
    \item One of three single-qubit gates $\rot{b} = \{\Id, \textsf{Y}_{-\!\frac{\pi}{2}}, \textsf{X}_{\frac{\pi}{2}}\}$ is applied prior to measurement, corresponding to measurement in the Pauli $z$-, $x$-, and $y$-bases.
\end{enumerate}
These steps are repeated for all combinations of initial state, channel duration, and measurement basis, and the results are saved in classical memory for analysis.
This dataset is fed to an MLE routine to determine the matrix elements of the density matrix $\rho_0$ and the positive operator-valued measures (POVMs) representing the imperfect measurement apparatus (Fig.~\ref{fig:LT_structure}b blue bubble, details in Sec.~\ref{subsec:spam}).
This initial characterization is then used to independently estimate the process map at each discrete time (Fig.~\ref{fig:LT_structure}b orange bubble and Sec.~\ref{subsec:process_map}) and the Hamiltonian and Lindblad operators for continuous time (Fig.~\ref{fig:LT_structure}b red bubble and Sec.~\ref{subsec:lindblad_extraction}).

Since Lindblad tomography is designed to characterize free evolution of a quantum system, note that a subset of the LT measurement sequences are identical to conventional $T_1$ (pale green gates in Fig.~\ref{fig:LT_structure}a) and $T_2$ (purple) measurements.  In this way, one can helpfully think of this measurement protocol as a hybrid of standard qubit characterization techniques, combining time-domain characterization of $T_1$/$T_2$ with process tomography.  By iterating over the full set of pre- and post-rotations, LT effectively pieces together all combinations of $T_1$- and $T_2$-like measurements to tomographically reconstruct the full quantum loss channel.

% --------------------------------------------------------------------------------------
\section{Extracting SPAM Errors} \label{subsec:spam}

Once we have collected the full set of data for the channel of interest, the subset of data obtained for the zero-duration channel $\tilde{\Id}(t\!=\!0)$ are analyzed to extract the SPAM errors for the device.

For the state-preparation errors, we parameterize the imperfect initial ground state of the qubit as an arbitrary single-qubit density matrix, $\rho_0$, which is constrained by physicality conditions to be positive semi-definite and have unit trace.  The condition of positivity is enforced in the optimization by expressing $\rho_0$ as a Cholesky decomposition $\rho_0=AA^{\dagger}$ and estimating the elements of the unconstrained matrix $A$ from which $\rho_0$ is computed. The unit trace condition is readily included by normalization.  

Measurement errors are characterized by extracting the positive operator-valued measure (POVM) describing the measurement apparatus. For most qubit modalities, measurements are natively performed in a fixed $z$-basis, while measurement in other bases are performed by rotating the state prior to measurement. 
% Conveniently, these rotations are a subset of the pulses needed to prepare the initial basis states of the Pauli matrices, and they thus do not require separate characterization. 
The single-qubit POVM corresponding to measurement in the $z$-basis has two operator elements ($2\times2$ matrices) $\{M_0,M_1\}$, where $M_0+M_1=\Id$. The probability of measuring a state $\rho$ in the ground state is then $p_0=\text{Tr}[\rho M_0]$ and the excited state is $p_1=\text{Tr}[\rho M_1]$. To estimate the POVM, we optimize over the matrix elements of $M_0$, subject to the constraint that both $M_0$ and $M_1=\Id-M_0$ are positive semidefinite.

In general, the fiducial gates required to initialize the cardinal states and rotate the measurement basis cannot be assumed error free. In order to fully characterize these operations, one would therefore parameterize these gates as arbitrary rotation matrices and estimate them together with the POVM and initial state parameters, in much the same way as in gate set tomography~\cite{Kim2015}. However, in Lindblad tomography, we significantly simplify the analysis by excluding the effects of imperfect rotation from our estimation. Here, our motivation is two-fold. First, we note that, for many NISQ-era devices across hardware modalities, errors due to imperfect measurement and ground state preparation exceed single-qubit gate errors. Second, since our ultimate goal is to characterize the idling channel over several multiples of the qubit's $T_1$ and $T_2$ times (tens of microseconds for superconducting qubits, in comparison to tens of nanoseconds to implement a single-qubit gate) the channel errors are naturally amplified relative to the errors in the fiducial gates, regardless of their intrinsic magnitude (in much the same way as in GST and RB). Furthermore, while ignoring the contribution of these errors typically introduces the issue of gauge freedom, we note that errors in the idling channel are first order gauge invariant (FOGI), and the contribution of errors in the fiducial gates can be safely ignored in this scenario~\cite{osti_1581878,2019APS..MARP35006B,Nielsen_inprep}. We also note that randomized benchmarking, which is not influenced by SPAM errors, can be performed prior to LT to obtain an independent estimate of the rotation pulse errors.

\begin{figure}[t!]
\includegraphics{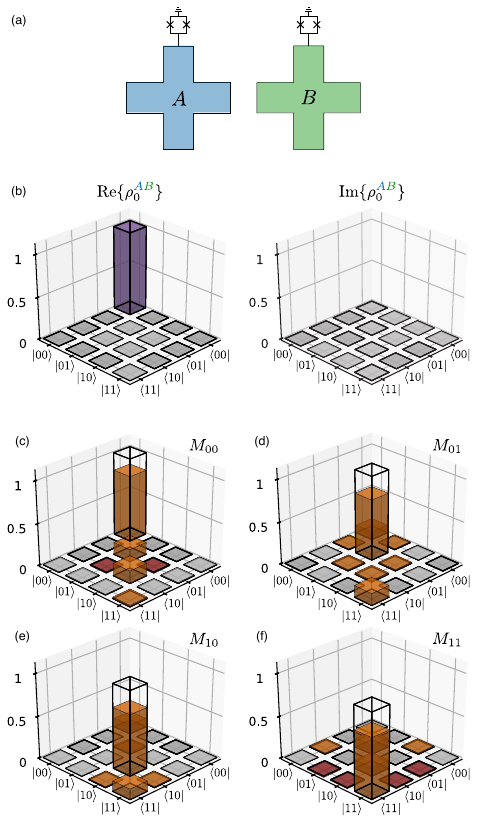}
\caption{Initial characterization of SPAM errors as part of Lindblad tomography.
	(a) Schematic of the two coupled transmon qubits used in this experiment.
	(b) Skyscraper plots of the imperfect two-qubit ground state $\rho_0^{AB}$ extracted during two-qubit LT
	% negative values shown in pale green, 
	(ideal ground state shown in wireframe, elements smaller than $10^{-2}$ omitted for visual clarity).
	(c--f) Extracted two-qubit POVMs, corresponding to imperfect measurement of the states $\ket{00}$, $\ket{01}$, $\ket{10}$, and $\ket{11}$ respectively (negative values shown in red, perfect POVMs shown in wireframe, imaginary parts and elements smaller than $10^{-2}$ omitted for visual clarity).  The full single- and two-qubit matrices for the extracted initial states and POVMs are included in Appendix~\ref{appendix:SPAM_lindblad_estimation}.
	}
	\label{fig:SPAM}
\end{figure}

In order to find the initial state $\rho_0$ and POVM $M_0$ that best describes the measurements, we construct a maximum likelihood function $\Lspam$ for our SPAM errors, which allows us to optimize over the unknown elements of these matrices. To perform a log-likelihood estimation, we take the logarithm of this function
\begin{align} \label{eq:loglike1}
\ln(\mathcal{L}_{\text{SPAM}})=\sum_{b,s}& f(s,b)\ln\Big(\text{Tr}\left[\rho_{s}M_{b}\right]\Big)\nonumber \\
&+\bar{f}(s,b)\ln\Big(\text{Tr}\left[\rho_{s}(\Id - M_{b})\right]\Big)
\end{align}
where $b\in\{z,x,y\}$ runs over the measurement bases, $s\in \{\ket{0},\ket{1},\ket{+},\ket{-},\ket{+i},\ket{-i}\}$ runs over the imperfect input states, and $f(s,b)$ ($\bar{f}(s,b)$) is the total number of `$0$'s (`$1$'s) recorded during repeated measurements of state $\rho_{s}$ in measurement basis $b$. In general, reconstructing the matrices $M_{b}=\rot{b}^{\dagger} M_0 \rot{b}$ would require estimating the matrix elements of both $M_0$ as well as the potentially faulty rotations ($\rot{b} = \{\Id, \mathsf{Y}_{-\frac{\pi}{2}}, \mathsf{X}_{\frac{\pi}{2}}\}$), and reconstructing the matrices $\rho_{s}=\rot{s} \rho_0 \rot{s}^{\dagger}$ would require estimating the matrix elements of both the initial state ($\rho_0$) and the potentially faulty rotations ($\rot{s} = \{\Id, \textsf{X}_{\pi}, \textsf{Y}_{\pm\!\frac{\pi}{2}}, \textsf{X}_{\mp\!\frac{\pi}{2}} \}$). However, as detailed above, LT strategically ignores the contribution of fiducial gate errors, so only the elements of $M_0$ and $\rho_0$ remain to be found.

The experimental data required for estimating SPAM errors corresponds to the subset of LT data taken for the zero-duration channel $\tilde{\Id}(t\!=\!0)$, where we prepare the qubit in each of the states $\rho_{s}$ and immediately measure in each basis $b$. Qubit measurements are recorded as single-shots, and the outcomes are labeled as either `$0$' or `$1$'.
 % using standard superconducting qubit measurement techniques~\cite{Krantz2019}. 
The initial state and POVMs are then estimated by maximizing the log-likelihood in Eq.~\eqref{eq:loglike1} with respect to the unknown matrix elements of the density matrix and measurement operators.

Now, in principle, we note that the SPAM parameters could alternatively be determined simultaneously with the Hamiltonian and Lindblad operations, using the entire set of measured time steps (i.e. not just $\tilde{\Id}(t\!=\!0)$). While this alternative method may work for some applications, obviating the need for a separate SPAM estimation, it has a major practical drawback: when equal weighting is given to all time steps, it is possible to end up in a local minimum which well fits the data at longer times (where channel errors dominate over SPAM) but fails to fit it at shorter times (where SPAM errors tend to dominate). To avoid this scenario and ensure accurate estimation of the SPAM errors, our SPAM estimation focuses only the data at $t=0$, where we expect these errors to dominate. Once we have an accurate estimate of those errors at small time, we then include them in the characterization of the channel at long times, as in Sec.~\ref{subsec:lindblad_extraction}.

This technique extends naturally to multi-qubit systems.  For two qubits A and B, we represent the initial state as a general two-qubit density matrix $\rho^{AB}_0 $, and we characterize the measurement apparatus using four $4\times4$ POVM matrices $\{M_{00},M_{01},M_{10},M_{11}\}$, corresponding to measurement of the states $\ket{00}$, $\ket{01}$, $\ket{10}$, and $\ket{11}$ respectively.  To determine the matrix elements of the initial state and the POVMs, we maximize a log-likelihood function analogous to Eq.~\eqref{eq:loglike1} containing four terms (corresponding to measurement of each of the four two-qubit computational states) and summing over the full set of two-qubit pre- and post-pulses (discussed further in Sec.~\ref{subsec:process_map}).

As a proof of principle demonstration of Lindblad tomography, we perform the full protocol on a device consisting of three capacitively-coupled flux-tunable transmon qubits, where we denote the two qubits characterized in this experiment as qubit A and B respectively (Fig.~\ref{fig:SPAM}a, full device characterization found in Appendix~\ref{appendix:device_infrastructure} and Ref.~\cite{Kjaergaard2020_PRX}). In Fig.~\ref{fig:SPAM}b, we plot the extracted matrix elements of the imperfect two-qubit ground states $\rho_0^{AB}$, extracted during two-qubit LT. In Fig.~\ref{fig:SPAM}c--f, we plot the elements of the two-qubit POVM matrices, corresponding to measurement of the four two-qubit computational states respectively. Bold wireframes in Fig.~\ref{fig:SPAM}b--f highlight ideal ground state preparation and perfect $z$-basis POVM matrices for comparison. 

To quantitatively motivate our choice to exclude fiducial gate errors from our analysis, we note that, for high fidelity superconducting qubits, single-qubit operations are typically orders of magnitude less prone to error (typical error rates $<0.05\%$) than measurements (typical error rates $1\%$)~\cite{kjaergaard_2020_stateofplay}. In such systems, it is therefore reasonable to assume that errors in single-qubit rotations have negligible impact on state-initialization and POVM estimation in comparison to imperfect thermalization and measurement error. In the device used for the present experiment, we find unoptimized measurement fidelities on the order of $90\%$, in comparison to single-qubit gate fidelities $\sim\!\! 99.99\%$ measured using interleaved randomized benchmarking (see Appendix~\ref{appendix:1qb_gates}). 
% As such, for the following characterization of this particular device, we will assume perfect rotation pulses, requiring only estimation of the $z$-basis POVM and initial state.

Looking at the most likely state-preparation errors (full matrices are reported in Appendix~\ref{appendix:SPAM_lindblad_estimation}), we note that the initial single-qubit states are very similar to a thermal state of the form $\rho_{\text{thermal}}=a\ket{0}\!\!\bra{0}+(1-a)\ket{1}\!\!\bra{1}$, $a\in[0,1]$. Minimizing the trace distance $D(\rho_{\text{thermal}},\rho_0)=|\rho_{\text{thermal}}-\rho_{0}|/2$ (where $|M|=\text{Tr}\left[\sqrt{M^{\dagger}M}\right]$) between the estimated initial states and a thermal state, with respect to the thermal parameter $a$, we find minimal trace distances $D(\rho_{\text{thermal}},\rho^{A}_0)=0.01$ ($D(\rho_{\text{thermal}},\rho^{B}_0)=0.04$) for $a=0.999$ ($a=0.998$). This result is thus consistent with the observation that the imperfect ground state of a superconducting qubit at finite temperature can be approximated as a thermal state of an anharmonic oscillator~\cite{Krantz2019}.  

Traditionally, the SPAM errors for a single qubit in a multi-qubit device would be found by characterizing the qubit while all its neighbors are left in their ground states. This is also the case for the above characterization in our device, though we note that this implies that the manipulation and readout of qubit B should have no effect on the POVM and initial state of qubit A. The extent to which this holds can be tested by comparing the estimated single-qubit POVMs and initial states for both qubits with the joint two-qubit POVM and initial state estimated using full two-qubit LT. The probability of measuring qubit A in state $x$ and qubit B in state $y$ is $\text{Tr}[\rho M_{xy}]$, where $\rho$ is the two-qubit density matrix, $M_{xy}$ is a two-qubit POVM, and $x,y\in\{0,1\}$. Once again, the likelihood function for the two-qubit system is then optimized under the constraint that the POVM elements sum to identity, and Cholesky decompositions can be used to ensure that they are positive semidefinite. All details of the estimated two-qubit POVM can be found in Appendix~\ref{appendix:SPAM_lindblad_estimation}.

The deviation between the estimated single- and two-qubit POVMs can be quantified as the trace distance $D(M_{xy},M_x\otimes M_y)$, where $M_x$ ($M_y$) are the estimated single qubit POVMs for qubit A (B). We find 
a trace distance of 0.04 for $xy=\{00,01,10\}$ and 0.05 for $xy=\{11\}$.   
 %To compare the single- and two-qubit POVMs, we maximize the trace distance $D(\Lambda(\rho),\Lambda'(\rho))$ between the output state of the two-qubit and single-qubit measurement channels, with respect to the input state $\rho$.  Performing this calculation, we find a maximum trace distance of $0.05$. 
%viewing the measurement procedure as a quantum channel, where we ``forget'' the measurement outcome. Specifically, we define the output state of the measurement channel to be
%\begin{equation}
%\Lambda(\rho)=\sum_{x,y=0,1}\mathcal{P}_{xy}\text{Tr}\left[\rho M_{xy}\right]  
%\end{equation}    
%where $\rho$ is the input state and $\mathcal{P}_{xy}$ is the projector on state $\ket{xy}$. We now compare the measurement channel $\Lambda$, corresponding to the estimated two-qubit POVM, with the measurement channel $\Lambda'$, corresponding to the product of the estimated single-qubit POVMs $M'_{xy}=M_x\otimes M_y$.
The trace distance between the estimated initial two-qubit state $\rho^{AB}_0$ and the product state  $\rho_0^A\otimes\rho_0^B$ constructed from the single-qubit estimates is $D(\rho_{0}^{AB},\rho_0^A\otimes\rho_0^B)=0.05$. Comparing the initial two-qubit state $\rho^{AB}_0$ to a product of single-qubit thermal states (see above), we find a trace distance of 0.01 for thermal populations $a=0.998$ (qubit A) and $a=0.994$ (qubit B). The full single- and two-qubit matrices for the extracted initial states and POVMs are reported in Appendix~\ref{appendix:SPAM_lindblad_estimation} for reference.

% --------------------------------------------------------------------------------------
\section{Reconstructing the Kraus Operators at Discrete Times} \label{subsec:process_map}

Once we have performed an initial characterization of the SPAM errors for our device, we can proceed to characterize the quantum channel of interest.  For superconducting devices, qubits are primarily subject to amplitude damping and dephasing noise over time~\cite{Krantz2019}, and the corresponding decay rates are traditionally characterized with simple $T_1$ and $T_2$ measurements respectively. However, to accurately model multi-qubit devices or develop tailored error correction techniques~\cite{Tuckett2019}, it is important to directly characterize the structure of these channels as well as how they depend on the state of neighboring qubits, details which are not readily obtainable from standard single-qubit $T_1$ and $T_2$ measurements. To obtain this information, we use LT to extract the Lindbladian of the channel, which requires process tomography over varying idling channel durations $\tilde{\Id}(t_i)$. 

Before estimating the Lindbladian of the channel for continuous time $t$, we can first separately extract the instantaneous evolution maps of the channel at each discrete time step $t_i$, which can then be used to check the validity of the time-independent Markovian model. Any quantum operation can be described by a set of Kraus operators such that the final state is related to the initial state as $\rho=\sum_j\mathcal{K}_j\rho_0\mathcal{K}^{\dagger}_j,$
where the Kraus operators satisfy $\sum_j\mathcal{K}^{\dagger}_j\mathcal{K}_j=\Id$ for a trace-preserving process. Note that the Kraus operators are only unique up to a unitary transformation: a quantum channel can be described by two different but equivalent sets of Kraus operators $\{\mathcal{K}_j\}$ and $\{\mathcal{K}'_k\}$, which will be related through a unitary matrix $U$ such that $\mathcal{K}_j=\sum_k U_{jk}\mathcal{K}_k'$.  In standard process tomography, one therefore often estimates a process matrix $\chi$ which is unique in a specified operator basis. Since the process matrix can be readily calculated from the Kraus operators and vice-versa, one can choose either description without loss of generality. In this work, we chose to estimate the Kraus operators, however we note that we also used the same MLE approach to estimate the process matrix but found a slower convergence of the optimization compared to the Kraus estimation.  We believe this is likely due to the unitary freedom in fixing the elements of the Kraus matrices. In what follows, we will therefore describe the estimation of the Kraus operators.

\begin{figure}[t!]
\includegraphics{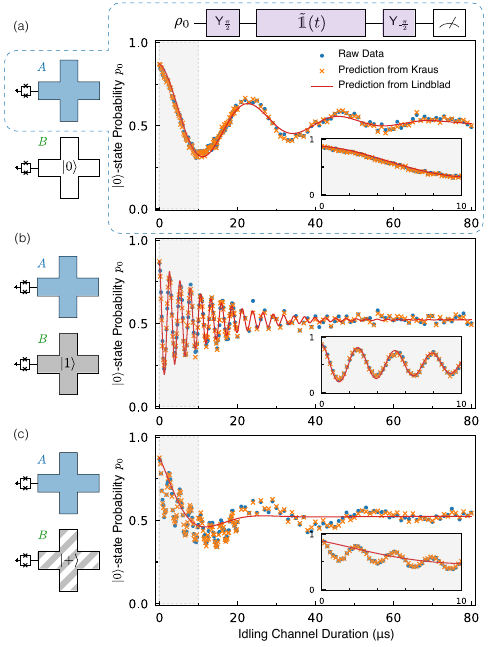}
\caption{Lindblad tomography applied to the idling channel of a single superconducting transmon qubit. (a) Data and analysis results for the LT sequence corresponding to a $T_2$ Ramsey measurement (purple gates in Fig.~\ref{fig:LT_structure}a), when the neighboring qubit is prepared near its ground state $\ket{0}$.  Blue points are $p_0$ of the state $\rho_{s}(t_i)$, averaged from 1000 single-shot measurements (discussed in Sec.~\ref{subsec:spam}, fitted value of $T_2$ recorded in the appendix, shot noise $1/\sqrt{N}\sim 3\%$). Orange $\times$'s are predicted measurement outcomes obtained from applying the Kraus operators estimated at each discrete time $t_i$ to the extracted initial state $\rho_0$ given an imperfect measurement $M_0$ (technique discussed in Sec.~\ref{subsec:process_map}). Red line traces the predicted outcomes for continuous time $t$, based on the most likely time-independent Lindblad and Hamiltonian operators (technique discussed in Sec.~\ref{subsec:lindblad_extraction}, average error = $2.25\times 10^{-2}$).  Results for the first 10$\mu$s are enlarged for clarity in inset.  We discuss the Lindblad fit for this particular dataset and its dependence on temporal fluctations during the protocol in Appendix~\ref{appendix:1qb_results}.  (b) The same measurement, taken when the neighboring qubit is near its excited state $\ket{1}$ (most likely Lindbladian in red, average error = $2.29\times 10^{-2}$).  The always-on $ZZ$-coupling between the two transmon qubits induces a state-dependent frequency shift when the neighbor is excited, which manifests here as a faster oscillation frequency.  (c) The same measurement, taken when the neighboring qubit is in a superposition state $\ket{+}$. In this basis, the always-on $ZZ$-coupling is an entangling operation, and the data is poorly predicted by the most likely single-qubit Lindbladian (red, average error = $6.91\times 10^{-2}$), a hallmark of non-Markovian evolution (see Sec.~\ref{sec:markvovianity}).}
\label{fig:1QB_full_LT}
\end{figure}

\begin{figure*}[t!]
\includegraphics{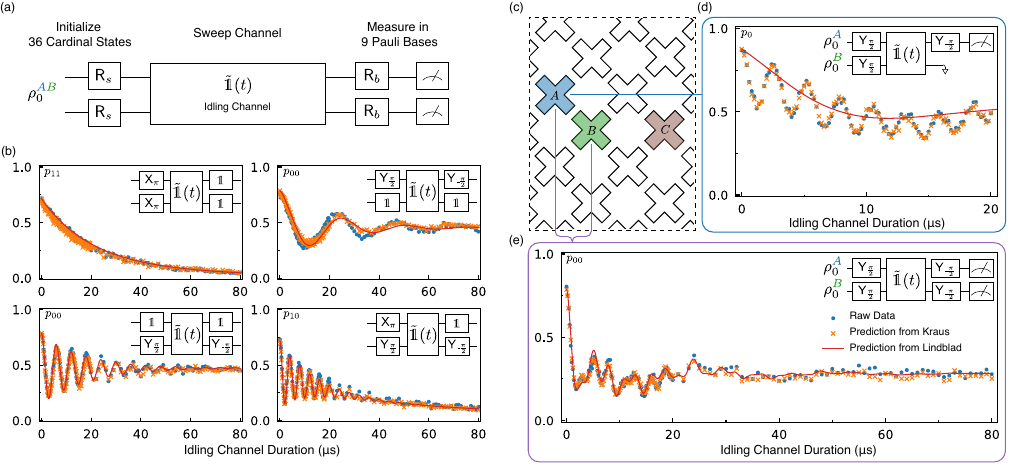}
\caption{Two-qubit Lindblad tomography protocol and results.  (a) Measurement protocol: the two qubits are initialized into their shared ground state $\rho_0^{AB}$ and prepared in each of 36 combinations of cardinal states; the channel of interest is swept; the qubits are rotated into each of nine combinations of Pauli bases and measured.  The full set of measurement results are passed through the same classical optimizer as in the single-qubit protocol, SPAM errors are extracted, and the instantaneous process maps and dynamic operators are estimated using MLE.  (b) Raw data (blue, shot noise $1/\sqrt{N}\sim 3\%$), predictions from extracted Kraus operators (orange), and predictions from estimated Hamiltonian and Lindblad operators (red) for several combinations of pre- and post-pulses. (c) Schematic of a large superconducting quantum processor, where the two qubits studied in this work are thought of as neighboring qubits (A and B) in a large patchwork. LT can be performed just as easily on distant qubits (i.e. A and C) to study non-local crosstalk.  (d) Single-qubit LT on qubit A while B is in a superposition state (same dataset as Fig.~\ref{fig:1QB_full_LT}c, enhanced for visual clarity).  The poor Lindblad fit (red, average error = $6.91\times 10^{-2}$) indicates that no single-qubit Lindblad operator successfully predicts the measured data; this observation, paired with the result of the Markovianity metric shown in Fig.~\ref{fig:figure5}c, suggests that the evolution is non-Markovian in the single-qubit frame.  (e) Two-qubit Lindblad tomography, where qubits A and B are both initialized in superposition states. While the pulse sequence is identical to (d), the data is now well-predicted by a two-qubit Lindbladian (average error = $2.15\times 10^{-2}$); this observation, paired with the result of the Markovianity metric shown in Fig.~\ref{fig:figure5}d, suggests that the channel is Markovian in the two-qubit frame.  Comparing (d) and (e) in concert with the results in Fig.~\ref{fig:figure5}, we conclude that the non-Markovian errors in the single-qubit data are due to spontaneous entanglement with qubit B (as discussed in Sec.~\ref{sec:markvovianity}), revealing the error source.}
\label{fig:figure4}
\end{figure*}

Characterizing the idling channel, the task is to estimate the Kraus operators describing the qubit evolution at discrete delay times. For each time delay $t_i \in [t_1,t_2,\ldots,t_N]$ we consider a maximum likelihood function of a similar form to Eq.~\eqref{eq:loglike1},
\begin{align} \label{eq:loglike2}
\ln(\mathcal{L}_\mathcal{K}(t_i))=\sum_{b,s} &f(s,b,i)\ln\big(\text{Tr}\left[\rho_{s}(t_i)M_{b}\right]\big)\nonumber \\
&+\bar{f}(s,b,i)\ln\big(\text{Tr}\left[\rho_{s}(t_i)(\Id - M_{b})\right]\big)
\end{align}
with parameters defined as in the SPAM estimation, except that $\rho_{s}(t_i)$ is now the discrete time evolution of the initial state $\rho_{s}$ at time $t_i$, under the evolution of the Kraus operators
\begin{equation} \label{eq:krausrhos}
\rho_{s}(t_i)=\sum_j\mathcal{K}_j(t_i)\rho_{s}\mathcal{K}^{\dagger}_j(t_i),
\end{equation}

The Kraus operators are then estimated by minimizing the log-likelihood function with respect to the unknown matrix elements of each Kraus operator, using the SPAM error parameters $\rho_{s}$ and $M_{b}$ found during the estimation in Section~\ref{subsec:spam}. For $N$ delay times, we obtain $N$ sets of Kraus operators where, for a $d$-dimensional quantum system, the process at each time is described by at most $d^2$ Kraus operators. Thus, for a single qubit, we estimate 4 Kraus operators per time delay (16 for two qubits).  

% Estimating the Kraus operators for each time step of the idling channel in our superconducting device, we start by initializing the minimization with the Kraus operators for single-qubit amplitude damping and dephasing noise. 
In Fig.~\ref{fig:1QB_full_LT}, we show the results of extracting the single-qubit Kraus operators for qubit A of our superconducting transmon device, superposed over the raw data obtained at each time step. Blue dots show the raw measurement probability $p_0$, averaged from 1000 single-shot measurements of the final state $\rho_{s}(t_i)$ (shot noise $1/\sqrt{N}\sim 3\%$), and orange $\times$'s show the predicted outcome of an imperfect measurement $M_b$ of the state $\rho_{s}(t_i)$, estimated by applying the extracted Kraus operators to the extracted imperfect initial state $\rho_{s}$ as in Eq.~\eqref{eq:krausrhos}. As such, the orange $\times$'s not only capture the channel noise, but also account for the SPAM errors of our device.  In Fig.~\ref{fig:1QB_full_LT}a--c, we compare results from the subset of LT sequences corresponding to a $T_2$-like measurement of qubit A (purple gates in Fig.~\ref{fig:LT_structure}a) when its nearest neighbor B is prepared close to either its $\ket{0}$-, $\ket{1}$-, or $\ket{+}$-state respectively.
% Comparing the raw data (blue) to our Kraus estimation (orange), we find that the extracted Kraus operators predict the measurement results for all three scenarios. However, 
Comparing these three scenarios, it is clear that the state of qubit B has a significant effect on the evolution of qubit A, a fact we examine in detail in Section~\ref{sec:markvovianity}. To capture the full dynamics of this interaction, it is thus necessary to extract the Kraus operators describing the full two-qubit channel.  

The estimation of the two-qubit Kraus operators follows from a straight-forward generalization of the single-qubit LT protocol, as shown in Fig.~\ref{fig:figure4}a.  The corresponding Kraus estimation is then performed by expanding the likelihood function in Eq.~\eqref{eq:loglike2} with all elements of the two-qubit POVM. A subset of the results of this extraction are shown in Fig.~\ref{fig:figure4}b,e. 
% As in the single-qubit case, we find that the estimated Kraus operators capture the dynamics remarkably well. 
Having obtained both the single- and two-qubit Kraus operators for this channel, we can investigate the Markovianity of the idling channel for our device.  In particular, we can directly investigate how Markovian two-qubit noise due to spurious interaction between qubits can manifest as non-Markovian single-qubit noise.

% --------------------------------------------------------------------------------------
\section{Validating The Markovian Model} \label{sec:markvovianity}

In this section, we use the Kraus operators extracted in the previous section to provide qualitative insight into whether or not the measured quantum channel can be fit to a Markovian model. There exist a number of proposed measures for non-Markovianity in the literature, and we refer the interested reader to reviews such as Refs.~\cite{Rivas2014,Vega2017} for reference. Notably, a number of experimental works have implemented the measure proposed in Ref.~\cite{Breuer2009}, which quantifies the backflow of information from the environment characteristic of non-Markovian error~\cite{Liu2011,Zelse2011}. This measure is also suitable for our purpose, because it considers the noise process over time, in contrast to instantaneous measures such as in Ref.~\cite{Wolf2008}. We note that other non-Markovianity measures exist that also consider the noise process over time and that may provide complementary information about the nature of non-Markovianity~\cite{Rivas2015}. However, for the purpose of simply assessing the validity of the Markovianity assumption of LT, the measure of  Ref.~\cite{Breuer2009} is sufficient.

The measure of Breuer et al. (Ref.~\cite{Breuer2009}) exploits the following fact: for any quantum process which can be captured by a time-dependent master equation of the form
\begin{align} \label{eq:mastereqn}
\dot{\rho}(t)=&-\frac{i}{\hbar}[\hat{H}(t),\rho(t)] \nonumber\\
&+\sum_i\gamma_i(t)\left(\hat{L}_i(t)\rho(t)\hat{L}^{\dagger}_i(t)
-\frac{1}{2}\{\hat{L}_i^{\dagger}(t)\hat{L}_i(t),\rho(t)\}\right)
\end{align}
with positive decay rates $\gamma_i(t)>0$, the trace distance $D(\rho_1(0),\rho_2(0))$ between two initial states $\rho_1(0),\rho_2(0)$ can only decrease. Here, $\hat{H}(t)$ and $\{\hat{L}_i(t)\}$ are the time-dependent Hamiltonian and jump operators of the process.

Since non-Markovian processes cannot be captured by a time-dependent master equation of the form in Eq.~\eqref{eq:mastereqn}, an increasing trace distance between two states under the evolution of a common channel signifies violation of Eq.~\eqref{eq:mastereqn} and thus the presence of non-Markovian errors.  Based on this observation, the measure $N_{\text{markov}}$ is suggested in Ref.~\cite{Breuer2009} as
\begin{equation} \label{eq:mmeasure}
N_{\text{markov}}=\text{max}_{\rho_1(0),\rho_2(0)}\int_{\sigma>0}\sigma(t,\rho_{1}(0),\rho_2(0))dt, 
\end{equation}
where $\sigma(t,\rho_{1}(0),\rho_2(0))=\frac{d}{dt}D(\rho_1(t),\rho_2(t))$. In other words, we integrate the derivative of the trace distance between a pair of states over all time intervals where the derivative is positive (i.e., trace distance increasing), and the larger the value of $N_{\text{markov}}$, the more non-Markovian the channel. However, we note that the quantitative value of $N_{\text{markov}}$ can be ambiguous, since the value is unbounded and extremely sensitive to experimental noise in individual data points. As such, rather than treat $N_{\text{markov}}$ as a quantitative metric, we instead treat it as a qualitative metric, plotting the trace distance as a function of the channel duration and looking for sustained periods of increasing trace distance. In Section~\ref{subsec:lindblad_extraction}, we complement this observation with a rigorous quantitative analysis of the error between the operator predictions and data.

\begin{figure}[t!]
\includegraphics{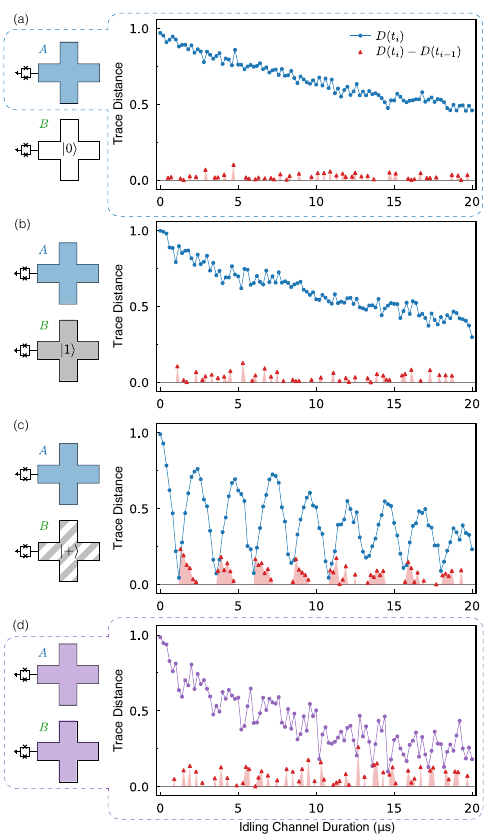}
\caption{Markovianity of single- and two-qubit idling channels.  (a--c) Qualitatively comparing the measured Markovianity of qubit A's idling channel when qubit B is prepared in $\ket{0}$, $\ket{1}$, or $\ket{+}$ respectively.  We find the two initial states of qubit A which together yield the largest value of $N_{\text{markov}}$, and we plot the trace distance $D$ between these two states at each time $t_i$ (blue points), as well as the difference in trace distance between sequential times (red triangles, values less than 0 omitted for visual clarity, since they do not contribute to $N_\text{markov}$).  Summing the area under the red points amounts to the discrete version of Eq.~\eqref{eq:mmeasure}, with sustained periods increasing trace distance indicating the presence of non-Markovian errors.
% , and $N_\text{markov}$ values are shown in each figure.  
When qubit B is prepared in $\ket{+}$ as in (c), we observe clear periods of increasing trace distance, suggesting the greatest presence of non-Markovian errors.  (d) The same protocol as above, taken using two-qubit LT and comparing the trace distance between two-qubit initial states (purple).  Maximizing over all two-qubit initial states, we find that none display the clear oscillations seen in (c), indicating the channel errors are largely Markovian in the two-qubit frame.}
\label{fig:figure5}
\end{figure}

Having estimated the Kraus operators for the single- and two-qubit idling channels of our coupled transmon system, we perform the optimization in Eq.~\eqref{eq:mmeasure} over the initial states of the LT protocol to calculate the measure $N_{\text{markov}}$. In Fig.~\ref{fig:figure5}, we use the results of the single- and two-qubit LT to perform this optimization, and we graphically illustrate $N_{\text{markov}}$ (given by the total area in red) for several qubit configuration. Notably, when qubit B is initialized in the state $\ket{+}$ (as in Fig.~\ref{fig:1QB_full_LT}c), the idling channel of qubit A registers clear periods of increasing trace distance (Fig.~\ref{fig:figure5}c). This behavior disappears when qubit B is initialized in either the $\ket{0}$- or $\ket{1}$-state, as well as in the combined two-qubit channel (Fig.~\ref{fig:figure5}a,b,d respectively). In these latter three scenarios, increases in the trace distance appear to arise from isolated statistical fluctuations in the data, with the trace distance otherwise monotonically decreasing over the channel duration.
% Furthermore, while the contributions to $N_{\text{markov}}$ in Fig.~\ref{fig:figure5}a,b,d appear to arise from individual statistical fluctuations in the data due to finite sampling, the increases in trace distance in Fig.~\ref{fig:figure5}c correspond to clear oscillations over the channel duration. 
% We note that, were one to interpolate the data, the impact of statistical fluctations in the distance measure would be significantly reduced, resulting in a smaller value of $N_{\text{markov}}$ for Fig.~\ref{fig:figure5}a,b,d, while the value for the case of Fig.~\ref{fig:figure5}c would be essentially unchanged.  This indicates that, when qubit B is prepared in $\ket{+}$, it induces distinctly non-Markovian errors in the noise environment of qubit A.

The distinctive presence of single-qubit non-Markovian behavior in Fig.~\ref{fig:1QB_full_LT}c is well understood from the physics of coupled transmon qubits. For two transmon qubits interacting via a fixed capacitance, the resulting dispersive repulsion of the $\ket{20}$- and $\ket{02}$-states shifts the frequency of the $\ket{11}$-state and gives rise to a ubiquitous ``always-on'' $ZZ$-interaction in the computational subspace of the form~\cite{Ganzhorn2020}:

\begin{equation} \label{eq:ZZcoupling}
\hat{H}_{zz}/\hbar = \omega_{zz} \ket{11}\!\!\bra{11} = \frac{\omega_{zz}}{4}(ZZ-ZI-IZ+II)
\end{equation}

\noindent where $\omega_{zz} = \omega_{11} - \omega_{01} - \omega_{10}$ is the energy shift of the $\ket{11}$-state due to the qubit coupling.

Consequently, when the two qubits are far detuned from each other, this interaction results in an effective two-qubit Hamiltonian of the form~\cite{OMalley2015}:

\begin{align}\label{eq:HwithZZ}
\hat{H}/\hbar=\omega_A \ket{10}\!\!\bra{10} &+ \omega_B \ket{01}\!\!\bra{01} \nonumber\\ 
&+ (\omega_A + \omega_B + \omega_{zz}) \ket{11}\!\!\bra{11}
\end{align}

\noindent where $\omega_A, \omega_B$ are the $\ket{0} \!\!\!\! \rightarrow \!\!\!\! \ket{1}$ transition frequencies of qubit A and B respectively. When one of the qubits is prepared in either $\ket{0}$ or $\ket{1}$, this interaction is manifest as a state-dependent frequency shift (hence the difference in oscillation frequency between Fig.~\ref{fig:1QB_full_LT}a and b).  However, when the two qubits are prepared in an initial state $\ket{++}$, they will evolve into an entangled state under the influence of the Hamiltonian in Eq.~\eqref{eq:HwithZZ}. If we only consider the evolution of one of the qubits---as is the case for the single-qubit Kraus estimation, where we are effectively tracing out one of the qubits---the always-on coupling will swap information between the qubit we are measuring and its neighbor. Qubit B thus functions as an environment with memory, and the entanglement between the two qubits gives rise to non-Markovian errors in the single-qubit picture. However, if the evolution of both qubits is considered, as in the two-qubit Kraus estimation, the always-on interaction is revealed to be unitary and this non-Markovian behavior disappears (Fig.~\ref{fig:figure5}d). 

% --------------------------------------------------------------------------------------
\section{Extracting the Lindbladian} \label{subsec:lindblad_extraction}
While we have shown how the Markovianity of the noise environment can be qualitatively assessed from the estimated Kraus operators, it is often difficult to extract much physical insight from the Kraus operators alone. If the channel is Markovian, one can apply LT to estimate the time-independent Lindbladian which best fits the measurement data for continuous times $t$ (right path in Fig.~\ref{fig:LT_structure}b). If, on the other hand, the channel is non-Markovian, one will be unable to find a set of operators which describe the data, since non-Markovian processes cannot be fit to a master equation. We note, however, that the Markovianity measure employed in Section~\ref{sec:markvovianity} only tells if the process can be captured by a general master equation with a time-dependent Lindbladian, and it does not guarantee that the assumption of a time-\emph{independent} Lindbladian is fulfilled. As such, comparison between the Markovinaity metric and the fit of the extracted Lindbladian allows us to qualitatively distinguish between three possibilities:
\begin{itemize}
\item The channel is Markovian and described by a time-independent Lindbladian. In this case, the extracted operators fit the data and the Markovianity measure will show a monotonically decreasing trace distance between pairs of states.
\item The channel is non-Markovian. In this case, the extracted operators poorly fit the data and the Markovianity measure will show clear periods of increasing trace distance.
\item The channel is Markovian but not described by a time-independent Lindbladian. In this case, the extracted operators poorly fit the data while the Markovianity measure shows a monotonically decreasing trace distance between pairs of states. The appearance of this phenomenon may also indicate failure in the MLE optimization itself, and additional analysis is required to confirm that the poor fit is physically meaningful.
\end{itemize}

For a time-independent Lindbladian, the master equation from Eq.~\eqref{eq:mastereqn} simplifies to  
\begin{equation} \label{eq:master2}
\dot{\rho}=-\frac{i}{\hbar}[\hat{H},\rho]+\sum_{i=1}^{d^2-1}\gamma_i(\hat{L}_i\rho\hat{L}^{\dagger}_i-\frac{1}{2}\{\hat{L}_i^{\dagger}\hat{L}_i,\rho\}).
\end{equation} 	
Choosing an operator basis $\{\sigma_i\}$ consisting of a Hilbert-Schmidt orthogonal set of traceless Hermitian operators in dimension $d$ (which can be constructed from tensor products of single-qubit Pauli matrices and the identity), the master equation can be rewritten as  
\begin{equation} \label{eq:master}
\dot{\rho}=-\frac{i}{\hbar}[\hat{H},\rho]+\sum_{i,j=1}^{d^2-1}L_{ij}(\sigma_i\rho\sigma^{\dagger}_j-\frac{1}{2}\{\sigma_j^{\dagger}\sigma_i,\rho\}),
\end{equation} 
where $L_{ij}$ is a Hermitian and positive semidefinite matrix capturing the incoherent evolution, commonly referred to as the Lindblad matrix. 

We note that, similar to the process matrix $\chi$, the Lindblad matrix is unique while the jump operators, like the Kraus operators, have a unitary freedom: the Lindblad equation in Eq.~\eqref{eq:master2} is invariant under a unitary transformation of the jump operators and decay rates. In particular, a new set of jump operators and decay rates $\{\sqrt{\gamma_i'}\hat{L}'_i\}$ can be constructed from the set $\{\sqrt{\gamma_j}\hat{L}_j\}$ as $\sqrt{\gamma_i'}\hat{L}'_i=\sum_j U_{ij}\sqrt{\gamma_j}\hat{L}_j$ where $U$ is a unitary matrix. Since the Lindblad matrix can readily be obtained from the decay rates and jump operators (and vice versa), one can choose either representation without loss of generality. In the following analysis, we choose to directly estimate the Lindblad matrix, and we derive the jump operators by diagonalizing this matrix. Specifically, a (unique) set of jump operators can be obtained by diagonalizing the Lindblad matrix as
\begin{equation}
\hat{L}_i=\sum_{j=1}^{d^2-1} U_{ij}\sigma_j
\end{equation}
where $U$ is a unitary matrix such that $L=U\mathds{D}U^{\dagger}$, where $\mathds{D}=\text{diag}(\gamma_1,\gamma_2,\ldots,\gamma_{d^2-1})$ is a diagonal matrix with the decay rates.

Performing LT, our goal is to estimate the Hamiltonian and the Lindblad matrix which most likely describes evolution under the idling channel for continuous time $t$.
% results of our single- and two-qubit measurements (blue data in Fig.~\ref{fig:1QB_full_LT} and Fig.~\ref{fig:figure4}). 
To do this, we must account for all combinations of initial states and measurement axes (as in our Kraus extraction), as well as for all channel durations. We therefore seek to maximize a log-likelihood function over all time steps $t_i$
\begin{equation}
\ln(\mathcal{L}_{\text{LT}})=\sum_{i=1}^N\ln(\mathcal{L}(t_i)),
\end{equation} 
where the likelihood function at each discrete time $\ln(\mathcal{L}(t_i))$ is defined as in Eq.~\eqref{eq:loglike2}, except that we no longer write $\rho_{s}(t_i)=\sum_i\mathcal{K}_i\rho_{s}\mathcal{K}^{\dagger}_i$. Instead, we find $\rho_{s}(t_i)$ by numerically solving the master equation in Eq.~\eqref{eq:master} for each guess at the elements of the Hamiltonian and the Lindblad matrices, where we evaluate the master equation at each time step $t_i$ by numerical exponentiation of the Lindbladian. As with the SPAM and Kraus estimation, a Cholesky decomposition is used to ensure that the Lindblad matrix is positive semidefinite. 

Once the most likely Lindbladian has been extracted, we evaluate the results of the optimization by calculating the error between the predictions of the operators and data. For each time step, the error for a given set of initial state and measurement axis is $|x^{\text{meas}}_i - x^{\text{model}}_i|$, where $x_i^{\text{meas}}$ is the measurement probability obtained in experiment at time step $t_i$ and $x_i^{\text{model}}$ is the corresponding estimate from the outcome of the MLE routine. The error is then averaged over time steps, and the average error for a given combination of initial state and measurement axis is reported. In addition, the $p$-value can be similarly calculated using a $\chi^2$ analysis of the extracted operators and data, as discussed in Appendix~\ref{appendix:error_analysis}.

We now turn to the results of the Lindblad estimation on our superconducting transmon device, and we consider a set of additional analyses motivated by our understanding of the physics of superconducting transmon qubits. These analyses, while not generically necessary for LT, are helpful for interpreting the results of the Lindblad estimation, and similar tests can be employed on other hardware platforms. In Fig.~\ref{fig:1QB_full_LT}, we plot the results of single-qubit Lindblad estimation on qubit A for three different preparations of qubit B.
% Running LT on our superconducting transmon device, the evolution under the single-qubit idling channel---as predicted by the estimated Hamiltonian and Lindblad matrix for Qubit A---is shown in Fig.~\ref{fig:1QB_full_LT}. 
Here, the solid red line traces out the predicted measurement results for the continuous evolution of qubit A over all times $t$, as predicted by the most likely Hamiltonian and Lindblad matrices.  When qubit B is in either its $\ket{0}$-state (Fig.~\ref{fig:1QB_full_LT}a) or $\ket{1}$-state (Fig.~\ref{fig:1QB_full_LT}b), we find that the Lindblad evolution of qubit A well fits the results of measurement (average error = $2.25\times 10^{-2}$, $2.29\times 10^{-2}$ respectively). %and the prediction of our extracted discrete processes (orange $\times$'s). 
However, when the neighboring qubit is prepared in the superposition state $\ket{+}$ (Fig.~\ref{fig:1QB_full_LT}c), we find that our estimation fails to find a combination of Hamiltonian and Lindbladian which well fit the data (average error = $6.91\times 10^{-2}$).  As discussed above, this is expected given the non-Markovian signature we previously found from the Kraus estimation (Fig.~\ref{fig:figure5}c), since non-Markovian processes cannot be captured by a master equation of the form in Eq.~\eqref{eq:master}.

Having estimated the most likely single-qubit Lindbladian, we proceed to estimate the most likely two-qubit Lindbladian describing the coupled system of qubits A and B. In Fig.~\ref{fig:figure4}, we plot a subset of the results of Hamiltonian and Lindblad extraction using two-qubit LT (red solid line), and we show that the estimated operators well fit both measurement (blue dots) and the prediction of our extracted Kraus operators (orange $\times$'s) for all combinations of initial state and measurement axis. In particular, we note a successful Lindblad fit for the sequence where both qubits are prepared in superposition states $\ket{++}$ (Fig.~\ref{fig:figure4}e, average error = $2.15\times 10^{-2}$), indicating that the non-Markovian errors observed in the corresponding single-qubit channel (Fig.~\ref{fig:1QB_full_LT}c,~\ref{fig:figure4}d, average error = $6.91\times 10^{-2}$) largely disappear in the two-qubit frame. Average errors and $p$-values for the full set of single- and two-qubit results (both the Linblad and the Kraus) are included in Appendix~\ref{appendix:error_analysis}.

From the two-qubit data, we estimate the following two-qubit Hamiltonian
% \begin{widetext}
\begin{equation}
\label{eq:Hextraction}
\hat{H}_{e}=\hbar\left(
\begin{smallmatrix}
-0.001& 0.008+0.024i & 0.004-0.003i& 0.001+0.015i\\
0.008-0.024i& -1.035 & 0.000+0.098i& -0.019+0.004i \\
0.004+0.003i& 0.000-0.098i & -0.258& -0.009+0.000i \\
0.001-0.015i& -0.019+0.004i & -0.009+0.000i& 1.323
\end{smallmatrix} \right) %\quad \quad
\end{equation}
% \end{widetext}
with angular frequencies in units of $2\pi\times \text{MHz}$, relative to the lab-frame of the driving pulses used for single-qubit rotations. Were these pulses chosen resonant with the qubit frequencies, we would expect the diagonal elements of the Hamiltonian to be zero, except for the $\ket{11}$-state, where the always-on $ZZ$-coupling shifts the energy.

From the Hamiltonian extracted in Eq.~\eqref{eq:Hextraction}, we see that there is a frequency detuning of $\Delta\omega_A/2\pi = 0.258/2\pi\,\,\text{MHz} = 41.1\text{kHz}$ for qubit A and $\Delta\omega_B/2\pi = 1.04/2\pi\,\,\text{MHz} = 165\text{kHz}$ for qubit B, which give rise to the oscillations seen in Fig.~\ref{fig:1QB_full_LT}a (qubit A) and Fig.~\ref{fig:figure4}b bottom left figure (qubit B). The effect of the $ZZ$-coupling is evident when qubit B is excited close to its $\ket{1}$-state (Fig.~\ref{fig:1QB_full_LT}b), and we consequently observe an increase in the frequency offset for qubit A. From the extracted Hamiltonian, we can estimate the frequency shift from the $ZZ$-coupling to be $\omega_{zz}/2\pi=(1.32-(-0.26-1.04))/2\pi \,\,\text{MHz} = 416 \text{kHz}$, which is consistent with independent estimation from device parameters (see Appendix~\ref{appendix:SPAM_lindblad_estimation}).

Having estimated the unitary part of the idling channel, we turn to the extracted two-qubit jump operators and decay rates corresponding to information loss during the channel, and the full expressions of the extracted jump-operators and decay rates are shown in Appendix~\ref{appendix:SPAM_lindblad_estimation}. Motivated by our understanding of the dominant error mechanisms for superconducting qubits, we can compare our extracted operators to single-qubit amplitude damping (at finite temperature) and dephasing noise (i.e. $T_1$ and $T_2$ processes respectively), which correspond to jump operators of the form 
\begin{align}
\hat{L}_{d,1}&\propto\sigma_z\otimes\mathbb{I} \label{eq:jump1} \\
\hat{L}_{d,2}&\propto\mathbb{I}\otimes\sigma_z  \\
\hat{L}_{-,1}&\propto\sigma_-\otimes\mathbb{I} \\
\hat{L}_{-,2}&\propto\mathbb{I}\otimes\sigma_-  \\
\hat{L}_{+,1}&\propto\sigma_+\otimes\mathbb{I}\ \\
\hat{L}_{+,2}&\propto\mathbb{I}\otimes\sigma_+ \label{eq:jump4}
\end{align} 
where the $\sigma_z$ operator corresponds to dephasing and the operators $\sigma_-\!=\!\ket{0}\!\!\bra{1}$ and $\sigma_+=\sigma_-^{\dagger}$ correspond to amplitude damping to a thermal state at finite temperature. 

To investigate whether traditional $T_1$ and $T_2$ models accurately describe the evolution of our qubits, we run a separate maximum likelihood optimization of the Lindbladian, this time constraining the jump operators to be of the form in Eq.~\eqref{eq:jump1}--\eqref{eq:jump4} and leaving only the rates ($\gamma$) and Hamiltonian ($\hat{H}$) as free parameters. We will refer to this as the \emph{restricted} optimization, while the previous optimization over general jump operators is referred to as the \emph{free} optimization. We compare the output of this restricted optimization by calculating the diamond norm distance between the two Liouvillian superoperators,
\begin{equation}\label{eq:delta_diamond}
\delta(t)=\norm{\Phi(\mathds{L}_{\text{free}},t)-\Phi(\mathds{L}_{\text{restricted}},t)}_\Diamond
\end{equation}
where $\mathds{L}_{\text{free}}$ ($\mathds{L}_{\text{restricted}}$) is the Liouvillian superoperator corresponding to the free (restricted) optimization, and $\Phi(\mathds{L},t)$ is the Choi-matrix representation of $e^{\mathds{L}t}$. For a diamond norm distance $\delta$, the minimum error probability when trying to distinguish between the two channels for each measurement shot is $(1-\delta/2)/2$~\cite{Benenti2010}. For $t\leq80$ $\mu$s, we find that $\delta(t)\leq0.2$; evaluating the asymptotic limit, we find that $\delta(\infty)=0.06$, indicating that the two evolutions result in similar steady states (Fig.~\ref{fig:diamond}).
%(for a plot of $\delta(t)$ over the measured channel duration, see supplementary material). 

From this analysis, we conclude that the extracted jump operators from the unrestricted optimization are largely consistent with single-qubit amplitude damping and dephasing channels, confirming that standard $T_1$ and $T_2$ models describe the data reasonably well. However, the deviation from the single-qubit model is significant and consistent with the observed always-on interaction between the qubits, which can lead to two qubit decay channels. Further investigation is necessary to pinpoint the physical mechanisms responsible for these errors, a promising direction for future work.

We also compare the steady state of the two-qubit Lindbladian found in the free optimization ($\rho^{AB}_\text{ss}$) to the initial two-qubit state $(\rho_0^{AB})$ from the SPAM estimation. Calculating the trace distance between these two states, we find a distance $D(\rho^{AB}_\text{ss},\rho_{0}^{AB})=0.09$, indicating a slight deviation between the two. This is unexpected, since the superconducting qubits are initialized by waiting many multiples of $T_1$---letting them relax to the steady state of the idling channel---and one would therefore have expected the steady state to be identical to the initial state. We note, however, that the deviation is relatively small and may originate from the fact that we only fit to data up to 80$\mu$s ($\sim 2T_1$ for qubits A and B, as shown in Table~\ref{table:parameters}) and the qubits have not fully relaxed.

% --------------------------------------------------------------------------------------
\section{Conclusion and Discussion}
In this work, we have proposed a technique for extracting the time-independent Hamiltonian, jump operators, and corresponding decay rates of an experimental quantum channel, which we refer to as Lindblad tomography (LT). Combining aspects of process tomography and time-domain $T_1$/$T_2$ measurement with Hamiltonian, Lindblad, and SPAM error estimation based on maximum likelihood (MLE), Lindblad tomography provides detailed information about the errors and noise environment of physical quantum devices and can be used to identify sources of qubit-qubit crosstalk.  

Applying Lindblad tomography to the characterization of a small superconducting device, we demonstrate how the results provided by LT also allow us to quantify crosstalk in qubit readout, frequency offsets in driving pulses, and the strength of always-on interactions between qubits, which can result in single-qubit non-Markovain noise.
 % which we quantified using the measure of Ref.~\cite{Breuer2009}. 
Furthermore, our characterization shows that the noise environment which best describes the data is consistent with single-qubit amplitude damping ($T_1$) and dephasing ($T_2$) channels to a large extent. 

While much of this proof-of-principal study focused on noise processes which arise due to the presence of neighboring qubits, we note that LT can be naturally applied to study a broad range of noise sources which impact the idling channel---such as coupling to coherent two-level systems (TLSs)~\cite{Klimov2018}, dephasing due to photons in readout resonators~\cite{Schuster2005}, and interaction with quasiparticles~\cite{Serniak2018} in superconducting systems---all of which may leave traces in the extracted Lindbladian and result in varying degrees of non-Markovian error.  Additionally, we believe that further investigation of changes in the noise environment over time~\cite{Burnett2019,Proctor2020} could offer a promising direction for future inquiry, and we are confident that LT will prove a valuable tool in future work towards suppressing these errors using either quantum control or error correction techniques.    

As noted in the introduction, we conclude by reiterating that the number of measurements required for Lindblad tomography scales exponentially with the number of qubits, as with standard process tomography or GST. Thus, full characterization of a large quantum processor using LT remains experimentally impractical.  However, since crosstalk between qubits is almost entirely 2-body, characterization of all combinations of two-qubit patches on a large quantum processor using LT will nevertheless provide valuable insights into the collective noise environment of the full processor, and these measurements can be used to bootstrap higher-order errors~\cite{Govia2020,Lilly2020}.  As the number of two-qubit patches only scales as $\sim\!\! N^2$ for an $N$-qubit device (regardless of hardware platform), characterization of direct two-qubit crosstalk with Lindblad tomography can therefore, in principle, be done efficiently. Furthermore, for devices where it is reasonable to assume that crosstalk is restricted to pairs of qubits within a certain maximal separation (as may well be the case for devices with equally spaced qubits, as in a lattice of superconducting qubits), the number of pairs to be characterized would only scale as $\mathcal{O}(N)$, though investigating the validity of this approach remains the subject of future work.  As research scales to larger and more complex systems and the possible sources of crosstalk and unintentional qubit entanglement inevitably increases, we are confident that repeated Lindblad tomography of single- and two-qubit patches will provide an important step towards modeling the dynamics of large-scale quantum processors.

During the final preparation of this manuscript, the authors became aware of a recent and separate theory work which proposes, implements, and numerically benchmarks the fitting of tomography data to quantum noise models, and we direct the interested reader to that manuscript for comparison~\cite{Onorati2021}.

\begin{figure*}[t!]
\includegraphics[width=\textwidth]{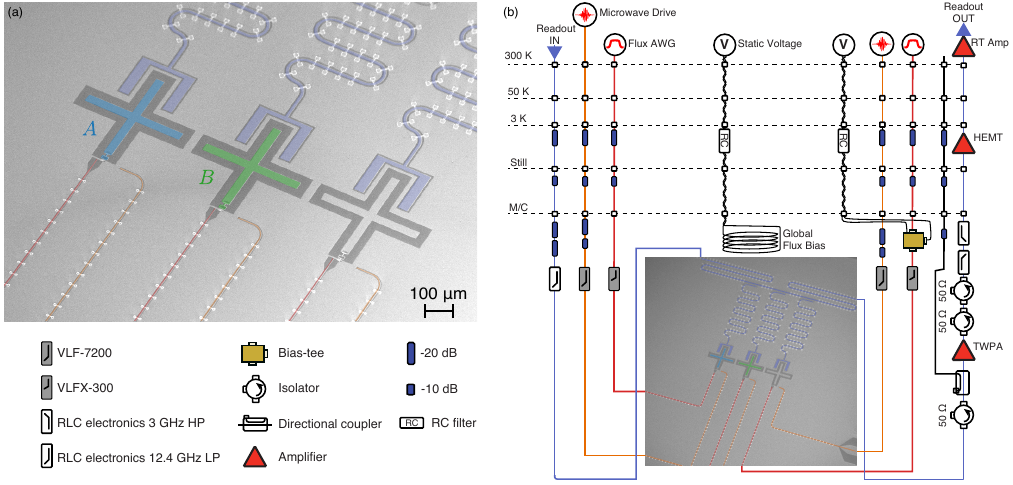}
\caption{Device and Wiring Diagram.  (a) SEM image of an identically fabricated copy of the device characterized in this work.  (b) Schematic of the control and readout hardware used to operate the quantum processor characterized in this experiment.  Dashed horizontal lines indicate the thermal stages of the dilution refrigerator in which the processor is measured.}
\label{fig:device_wiring}
\end{figure*}

\section*{Software and Data Availability}

The code used for analyzing the results of Lindblad tomography throughout this work can be found in the public GitHub repository~\cite{LTgithub}. The experimental data used in this work may be made available upon reasonable request of the corresponding authors and with the permission of the US Government sponsors who funded the work.

\begin{acknowledgments}
The authors gratefully acknowledge Agustin Di Paolo, Daniel Stilck França, and Bharath Kannan for valuable discussion and feedback on the manuscript;
Robin Blume-Kohout and the Quantum Performance Lab (QPL) at Sandia for substantial comments and discussion regarding the relationship between LT and GST; 
Chihiro Watanabe and Mirabella Pulido for their extensive assistance during this project;
and the MIT facilities and custodial staff for all their work in maintenance of the laboratory space.

AG acknowledges funding from the 2019 Google US/Canada PhD Fellowship in Quantum Computing.
JB acknowledges funding from the NWO Gravitation Program Quantum Software Consortium.
MC acknowledges financial support from the European Research Council (ERC GrantAgreement No. 81876), VILLUM FONDEN via the QMATH Centre of Excellence (Grant No.10059) and the QuantERA ERA-NET Cofund in Quantum Technologies implemented within the European Union’s Horizon 2020 Programme (QuantAlgo project) via the Innovation Fund Denmark.
MC also acknowledges the hospitality of the Center for Theoretical Physics at MIT, where part of this work was carried out.
MK was supported by Villum Fonden (grant 37467) through a Villum Young Investigator grant for part of this work.

This research was funded in part by the U.S. Army Research Office Grant W911NF-18-1-0411 and the Assistant Secretary of Defense for Research \& Engineering under Air Force Contract No. FA8721-05-C-0002. Opinions, interpretations, conclusions, and recommendations are those of the authors and are not necessarily endorsed by the United States Government.

\end{acknowledgments}

\appendix

\section{Device and Measurement Infrastructure}\label{appendix:device_infrastructure}

The quantum processor characterized in this work consists of three capacitively coupled superconducting flux-tunable transmon qubits arranged in a linear chain (Fig.~\ref{fig:device_wiring}a).  For this initial proof-of-principle demonstration of single- and two-qubit Lindblad Tomography, we chose to consider only the left and middle qubits of the chain, which we label qubit A and B respectively.  The rightmost qubit is far detuned to its frequency minimum and left to idle in its ground state for the duration of the characterization protocol.  Significant device parameters for qubits A and B are noted in Table \ref{table:parameters}.  In Fig.~\ref{fig:device_wiring}b, we outline the control and readout hardware used to perform gate operations and measure the state of the qubits inside a dilution refrigerator.  Additional characterization of the device used in this experiment can be found in Ref.~\cite{Kjaergaard2020_PRX}.

\begin{table}[t!]
\centering
 \begin{tabular}{|c|c|c|} 
 \hline
 Parameter & Qubit A & Qubit B \\ [0.5ex] 
 \hline\hline
 Idling Frequency, $\omega_i/2\pi$ & 4.744GHz & 4.222GHz \\ 
 Anharmonicity, $\eta/2\pi$ & -175MHz & -190MHz \\
 Coupling Strength, $g/2\pi$ & \multicolumn{2}{c|}{12MHz} \\
 Junction Asymmetry & 1:5 & 1:10 \\
 Single-qubit Gate Time & 30ns & 30ns \\ 
 Readout Resonator Frequency, $\omega_r/2\pi$ & 7.252GHz & 7.285GHz \\ 
 \hline\hline
 Energy Relaxation Time, $T_1$ &  26$\mu$s &  35$\mu$s \\
 Ramsey Decay Time, $T_2$ & 25$\mu$s & 24$\mu$s \\ %[1ex]
 \hline
 \end{tabular}
\caption{Device parameters for the two qubits characterized in this work.  Reported values of $T_1$ and $T_2$ are found by fitting the raw data from the corresponding subsets of the single-qubit LT sequence (highlighted gates in Fig.~1a of the main text, no pulses applied to the neighboring qubits) and recording the decay time of the fit, consistent with standard experimental convention.  LT generalizes this technique by extracting the decay channel from the full set of initial states, measurement axes, and channel durations, as emphasized in Fig.~\ref{fig:figurelindblad1}.}
\label{table:parameters}
\end{table}

\section{Single-Qubit Gate Characterization}\label{appendix:1qb_gates}

% In its most general formulation, Lindblad Tomography characterizes not only errors which arise during the channel of interest, but also errors in qubit state preparation and measurement (SPAM).

% However, since these three categories of error do not necessarily have independent experimental signatures, it is not always possible to fully disambiguate the relative contribution of each source of error.  
% To simplify our characterization protocol significantly, we add in the additional assumption that our state preparation and measurement axis rotations (\rot{s} and \rot{b}) have negligible error in comparison to thermalization, measurement, and channel errors.

\begin{figure*}[t!]
\includegraphics[width=\textwidth]{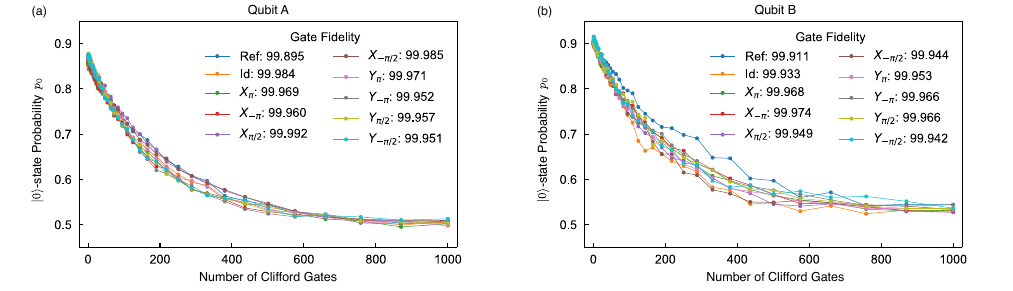}
\caption{Single-qubit Reference and Interleaved Clifford Randomized Benchmarking.  Characterization of nine single-qubit gates performed on qubit A (a) and qubit B (b).  Clifford reference fidelity (Ref) and interleaved gate fidelities are recorded in the legend.  This set includes all the gates required for state preparation (\rot{s}) and measurement axis rotation (\rot{b}).  Note that each of these gates exceeds an interleaved RB fidelity of 99.9\%, above the threshold for discounting rotation errors in our Lindblad Tomography protocol.}
\label{fig:RB_supp}
\end{figure*}

As we note in the main text, Lindblad tomography is resilient to errors in the single-qubit fiducial gates \rot{s} and \rot{b} required required for state preparation and measurement axis rotation. Nonetheless, we can independently characterize these rotations by performing Interleaved Clifford Randomized Benchmarking (IRB) on the full set of single-qubit gates $\textsf{R}\in \{ \Id, \textsf{X}_{\pi}, \textsf{Y}_{\pm\!\frac{\pi}{2}}, \textsf{X}_{\mp\!\frac{\pi}{2}} \}$ required to run Lindblad Tomography on each qubit.  For each of these operations, we record fidelities in excess of 99.9\%, over an order of magnitude greater than the fidelity observed for state initialization or measurement (Fig.~\ref{fig:RB_supp}).

\section{SPAM and Lindblad Estimation}\label{appendix:SPAM_lindblad_estimation}
We estimated the SPAM errors both for each qubit individually and for the combined two-qubit system. In both cases, the maximization of the log-likelihood function was performed using MATLAB's built-in function \texttt{fmincon}, which is a gradient-based numerical optimizer for constrained nonlinear problems. For documentation please see Ref.~\cite{bib_fmincon}. The function was run with the interior-point algorithm. We note that, to avoid the optimizer getting stuck in local minima, multiple starting points were tried to find a good approximation of the global minimum. This was done by sampling random POVMs and initial states as starting points. For both estimations we sampled over $10^4$ different starting points. We ran optimizations where we restricted the search space to be within some deviation of perfect POVMs and zero temperature thermal initial state for varying deviations. In particular, we restricted the initial thermal population of the $\ket{1}$-state to be smaller than 5\%, a choice which was motivated by estimation of the effective device temperature for the two transmon qubits in a dilution refrigerator at a base temperature of 11mK~\cite{Jin2015}.   %As discussed in the main text, the size of problem scales exponentially with the number of qubits and thus the performed optimization is not suited for a many-qubit SPAM estimation. 
For the single-qubit estimation of qubit A, we used data where the neighboring qubit B was kept in the ground state and measured in the $z$-basis, such that no pulses were applied to qubit B during the measurement (and vice versa for estimation of qubit B. The POVMs found from the maximization are
\begin{align}
M^{A}_{0}&=\left(
\begin{smallmatrix}
0.870& 0.00+0.015i \\
0.00-0.015i& 0.168
\end{smallmatrix} \right), \\
M^{B}_{0}&=\left(
\begin{smallmatrix}
0.880& -0.004-0.031i \\
-0.004+0.031i& 0.165
\end{smallmatrix} \right)\label{eq:POVM1}
\end{align}    
indicating that there are significant measurement errors on the order of 10-20\%. The estimated initial states are
\begin{align}
\rho^{A}_{0}&=\left(
\begin{smallmatrix}
0.999& -0.002-0.005i \\
-0.002+0.005i& 0.001
\end{smallmatrix} \right), \\
\rho^{B}_{0}&=\left(
\begin{smallmatrix}
0.998& 0.009-0.04i \\
0.009+0.04i& 0.002
\end{smallmatrix} \right)\label{eq:rhoSPAM1}
\end{align} 
which are consistent with single qubit thermal states as discussed in the main text. The maximized log-likelihood was $\ln(\mathcal{L}_{\text{SPAM}})=-1.083\times10^4$.

For the two-qubit system, we found the following initial state and POVM elements
\begin{widetext}
\begin{align}
\rho^{AB}_{0}&=\left(
\begin{smallmatrix}
0.992& -0.001+0.000i & 0.000+0.000i & 0.000-0.001i \\
-0.001+0.000i& 0.004 & 0.001-0.002i & 0.000+0.003i \\
0.000+0.000i& 0.001+0.002i & 0.001 & -0.001+0.000i \\
0.000+0.001i& -0.003i & -0.001+0.000i & 0.003
\end{smallmatrix} \right),\\
M_{00}&=\left(
\begin{smallmatrix}
0.7920& -0.010-0.019i & -0.008+0.010i & 0.006-0.002i \\
-0.010+0.019i& 0.146 & -0.014+0.000i & 0.010-0.004i \\
-0.080-0.010i& -0.014+0.000i & 0.151 & 0.006-0.007i \\
0.006+0.002i& -0.010+0.004i & 0.006+0.007i & 0.018
\end{smallmatrix} \right), \\
M_{01}&=\left(
\begin{smallmatrix}
0.095& 0.010+0.014i & 0.002+0.002i& -0.009+0.002i \\
0.010-0.014i& 0.726 & 0.014+0.002i & 0.003+0.004i \\
0.002-0.002i& 0.014-0.002i & 0.018 & -0.004-0.002i \\
-0.009-0.002i& 0.003-0.004i & -0.004+0.002i & 0.130 
\end{smallmatrix} \right), \\ 
M_{10}&=\left(
\begin{smallmatrix}
0.110& 0.00+0.002i & -0.008-0.009i & 0.002-0.003i \\
0.000-0.002i& 0.017 & 0.006-0.010i & 0.002-0.001i \\
-0.008+0.009i& 0.006+0.010i & 0.727 & 0.011-0.001i \\
0.002+0.003i& 0.002+0.001i & 0.011+0.001i & 0.124 
\end{smallmatrix} \right), \\ 
M_{11}&=\left(
\begin{smallmatrix}
0.002& 0.000+0.003i & 0.014-0.003i & 0.001+0.003i \\
0.000-0.003i& 0.111 & -0.005+0.007i & -0.015+0.001i \\
0.013+0.003i& -0.005-0.007i & 0.104 & -0.013+0.010i \\
0.001-0.003i& -0.015-0.001i & -0.013-0.010i & 0.718 
\end{smallmatrix} \right).
\end{align}
\end{widetext}
% \newpage
% and an initial two-qubit state of
% \begin{widetext}
% \begin{equation}
% \rho^{AB}_{0}=\left(
% \begin{smallmatrix}
% 0.992& -0.001+0.000i & 0.000+0.000i & 0.000-0.001i \\
% -0.001+0.000i& 0.004 & 0.001-0.002i & 0.000+0.003i \\
% 0.000+0.000i& 0.001+0.002i & 0.001 & -0.001+0.000i \\
% 0.000+0.001i& -0.003i & -0.001+0.000i & 0.003
% \end{smallmatrix} \right).
% \end{equation}
% \end{widetext}

As discussed in the main text, these are consistent with the estimated single qubit POVMs and initial states indicating that there is not much cross talk between the qubits in their initial state and the measurement operation. The maximized log-likelihood was $\ln(\mathcal{L}_{\text{SPAM}})=-3.868\cdot10^5$.

For the Kraus and Lindblad optimizations we maximized the log-likelihood function using MATLAB's built-in functions \texttt{fmincon} and \texttt{fminsearch}. For the Kraus optimization, \texttt{fmincon} was used in for both the single-qubit and two-qubit data in order to enforce the constraint of a trace-preserving map. As with the SPAM characterization, the interior-point algorithm was used~\cite{bib_fmincon}. As an initial point of the optimization, we used Kraus operators corresponding to an Identity channel to estimate the Kraus operators of the first time step, where the evolution from the initial state is assumably small. The resulting estimate was then used as starting point for the next time step and this procedure was iterated for the whole time series.  

For the Lindblad estimation, all physically constraints of the evolution could be ensured by employing a Cholesky decomposition of the Lindblad matrix and using the Hermiticity of the Hamiltonian to reduce the number of free parameters. The optimization therefore allowed for an unconstrained optimizer and we used the \texttt{fminsearch} optimizer of MATLAB. For documentation, please see Ref.~\cite{matlab2}. This algorithm uses the simplex search method of Ref.~\cite{lagarias1998}, which is a gradient free method.  As initial starting point for the optimization, we used the Lindblad matrix corresponding to pure dephasing and (zero-temperature) amplitude damping for the single qubit Lindblad estimation. For the two-qubit optimization we used the estimated single-qubit Hamiltonian and Lindblad operators as initial point.     %As discussed in the main text, the size of problem scales exponentially with the number of qubits and thus the performed optimization is not suited for a many-qubit Lindblad estimation.
The Lindblad operators estimated for the two-qubit system are shown below.
\begin{widetext}
\begin{eqnarray}\label{eq:Lindblad2} 
\hat{L}_{1}&=&\left(
\begin{smallmatrix}
0.501+0.001i& 0.000+0.001i & -0.0002+0.002i & -0.002+0.000i \\
0.000-0.001i& 0.499-0.001i & -0.001+0.001i & -0.002+0.002i \\
-0.001-0.001i& 0.001-0.001i & -0.499+0.000i & 0.000+0.001i \\
0.002+0.000i& -0.001-0.001i & 0.000+0.000i & -0.501+0.000i 
\end{smallmatrix} \right), \\
\hat{L}_{2}&=&\left(
\begin{smallmatrix}
0.448-0.001i& 0.109+0.246i & 0.001-0.002i & 0.002-0.002i \\
0.061-0.125i& -0.451+0.002i & -0.001+0.002i & 0.002+0.005i \\
-0.003-0.002i& -0.003+0.001i & 0.454+0.001i & 0.109+0.249i \\
-0.004+0.001i& 0.000+0.002i & 0.064-0.131i & -0.451-0.002i 
\end{smallmatrix} \right), \\
\hat{L}_{3}&=&\left(
\begin{smallmatrix}
-0.072+0.161i& 0.639-0.031i & -0.003-0.001i & 0.007+0.00i \\
0.114+0.039i& 0.072-0.161i & -0.001+0.001i & -0.005-0.001i \\
0.002+0.003i& 0.001+0.002i & -0.071+0.162i & 0.655-0.044i \\
0.002-0.003i& -0.001-0.002i & 0.134+0.035i & 0.071-0.163i
\end{smallmatrix} \right), \\
\hat{L}_{4}&=&\left(
\begin{smallmatrix}
0.004+0.000i& -0.001-0.003i & 0.000-0.703i & -0.003+0.00i \\
0.000-0.001i& 0.000-0.002i & -0.004-0.002i & 0.000-0.703i \\
0.001+0.078i& 0.002+0.001i & 0.000+0.002i & -0.002-0.003i \\
0.000-0.001i& -0.001+0.079i & 0.000-0.001i & -0.004+0.000i 
\end{smallmatrix} \right).
\end{eqnarray}
\end{widetext}
The corresponding decay rates were found to be $\gamma_1=0.071$ MHz, $\gamma_2=0.097$ MHz, $\gamma_3=0.042$ MHz, and $\gamma_4=0.055$ MHz. We fixed the unitary freedom of the jump operators by deriving them from the diagonalization of the Lindblad matrix, as discussed in the main text, and we normalize them such that $\text{Tr}\{\hat{L}\hat{L}^{\dagger}\}=1$. The extracted Hamiltonian is shown in Eq.~\eqref{eq:Hextraction} in the main text.

\begin{figure}[t!]
\includegraphics[width=.5\textwidth]{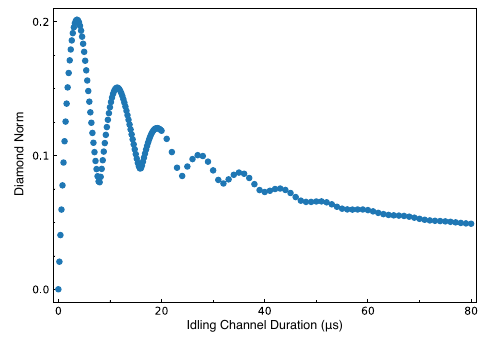}
\caption{Deviation between the estimated Liouvillian of the restricted and free optimization. The deviation is calculated as the diamond norm of the difference between the two superoperators, as in Eq.~\eqref{eq:delta_diamond}.}%{}
\label{fig:diamond}
\end{figure}

\begin{figure*}[t!]
\includegraphics[width=\textwidth]{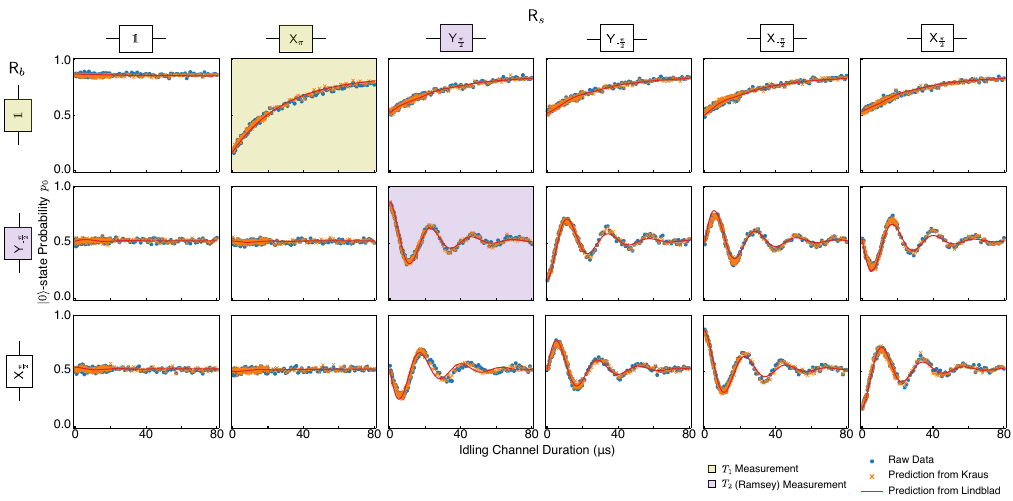}
\caption{Single-qubit LT results for the full set of initial states and measurement bases, performed on qubit A while qubit B is its ground state. The purple $T_2$ plot is the same used in Fig.~\ref{fig:1QB_full_LT}a in the main text. We emphasize that, while standard $T_1$ and $T_2$ metrics are determined by simply fitting the measurement results of a single set of pre- and post-pulses (pale green or purple, respectively), the Kraus operators (orange) and Lindbladian (red) here are determined from the full set of 18 measurement sequences (36 initial states $\times$ 9 measurement bases $= 324$ sequences for two-qubit LT).}
\label{fig:figurelindblad1}
\end{figure*}

As noted in the main text, the jump operators above are similar to single-qubit amplitude damping and dephasing channels, and we refer to the main text for comparison of the extracted operators to this model. For this comparison, we also ran a restricted optimization with the jump operators fixed to correspond to single-qubit dephasing and finite temperature amplitude damping noise. For this restricted optimization, we extracted a Hamiltonian
% \begin{widetext}
\begin{equation}
\hat{H}=\hbar\left(
\begin{smallmatrix}
0.001& 0.001+0.02i & 0.003-0.002i & 0.009-0.012i\\
0.001-0.020i& -1.031 & 0.007+0.003i& -0.017-0.001i \\
0.003+0.002i & 0.007-0.003i & -0.257& -0.012+0.000i \\
0.009+0.012i& -0.017+0.001i & -0.012+0.000i& 1.328
\end{smallmatrix} \right)
\end{equation}
% \end{widetext}
with angular frequencies in units of $2\pi\times \text{MHz}$ and decay rates of $\gamma_{1,d}=0.071$ MHz,  $\gamma_{2,d}=0.091$ MHz, $\gamma_{-,1}=0.055$ MHz,$\gamma_{-,2}=0.048$ MHz, $\gamma_{+,1}=0.001$ MHz, and $\gamma_{+,2}=0.004$ MHz. The deviation between the restricted and free optimization, $
\delta(t)$, as defined in the main text, is seen in Fig.~\ref{fig:diamond}. We note that the maximum likelihood was found for the unrestricted optimization ($\ln(\mathcal{L}_{\text{LT}})=-7.121\cdot10^7$ for the unrestricted compared to $\ln(\mathcal{L}_{\text{LT}})=-7.124\cdot10^7$ for the restricted optimization). 

Extracting the unrestricted Hamiltonian for the two-qubit system, we find an estimated state-dependent frequency shift $\omega_{zz}/2\pi = 416 \text{kHz}$ due to the always-on $ZZ$-coupling.  As a check, the frequency shift $\omega_{zz}$ can be independently estimated from device parameters using the relation~\cite{OMalley2015}:

\begin{equation}
\omega_{zz} = \frac{2g^2}{\Delta - \eta_B} + \frac{2g^2}{-\Delta - \eta_A}
\end{equation}

\noindent where $\eta_A, \eta_B$ are the anharmonicity of qubit A and B respectively, $g$ is the coupling strength between the two qubits, and $\Delta = \omega^A_i-\omega^B_i$ is the frequency detuning between them.  Substituting in the parameters for our device from Table~\ref{table:parameters}, we estimate a state-dependent frequency shift $\omega_{zz}/2\pi = 425\text{kHz}$, consistent with the value found from the Hamiltonian extraction using LT.

\section{Single-Qubit LT Results and Discussion}\label{appendix:1qb_results}

The estimated jump operators for the single-qubit noise channel of qubit A with the neighboring qubit in the ground state are
\begin{align} %\label{eq:jump1}
\hat{L}_{1}&=\left(
\begin{smallmatrix}
-0.551-0.052i& 0.030-0.622i \\
0.030-0.010i& 0.551+0.052i
\end{smallmatrix} \right), \\
\hat{L}_{2}&=\left(
\begin{smallmatrix}
0.438-0.019i& 0.144+0.757i \\
0.144-0.042i& -0.438+0.019i
\end{smallmatrix} \right)
\end{align}    
with decay rates $\gamma_1=0.029$ MHz and $\gamma_2=0.037$ MHz. 
We note that these operators fit the data well, as shown in Fig.~\ref{fig:figurelindblad1}, 
% and are similar to single-qubit dephasing ($\hat{L}_1$) and finite temperature amplitude damping ($\hat{L}_2$) channels up to a unitary transformation. 
and correspond to a maximized log-likelihood of $\ln(\mathcal{L}_{\text{LT}})=-1.739\cdot10^6$. 

Running LT on the full set of pre- and post-pulses for this particular dataset, we note that there is some small disagreement between the Lindblad fits (red) and the data (blue) for sequences corresponding to $T_2$-type measurements.  For example, we observe that the Lindblad fit slightly underestimates the decay time for the dataset when qubit A is prepared in $\ket{+}$ and measured in the $x$-basis (Fig.~\ref{fig:1QB_full_LT}a in the main text, purple highlighted plot in Fig.~\ref{fig:figurelindblad1}).  Looking at the full matrix of pre- and post-pulses, we note small temporal fluctuations in the channel over the course of the data aquisition period, and we see that the Lindblad fit consequently overestimates the decay time of some traces relative to others.  Since LT finds the single time-independent Lindblad operator which best describes \emph{all} combinations of pre- and post-pulses, spurious temporal fluctuation in the channel during a small set of measurements constitutes a partial violation of Assumption 1 of LT%~\cite{maintext}
, and this fluctuation will affect the fit of the other datasets.  

% To confirm this, we rerun our Lindblad estimation on this dataset, first excluding the single dataset corresponding to preparation of the $\ket{-i}$-state and measurement in the $y$-basis (Fig.~\ref{fig:figurelindblad1_excl2}), and then excluding all datasets where qubit A is prepared in $\ket{-i}$ (Fig.~\ref{fig:figurelindblad1_excl1}).  Excluding these datasets from the fit, we see that we find a Lindblad operator which even better matches the remaining datasets.

\section{Error and $\chi^2$ Analysis}\label{appendix:error_analysis}

\begin{figure*}[t!]
	\centering
	\includegraphics[scale=1]{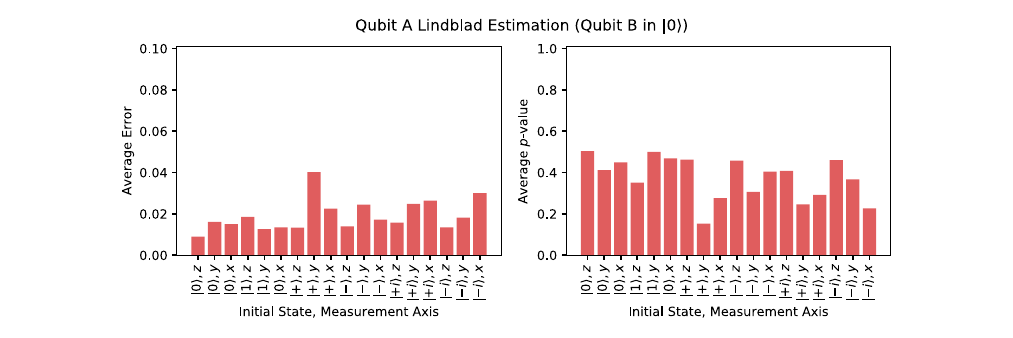}
	\includegraphics[scale=1]{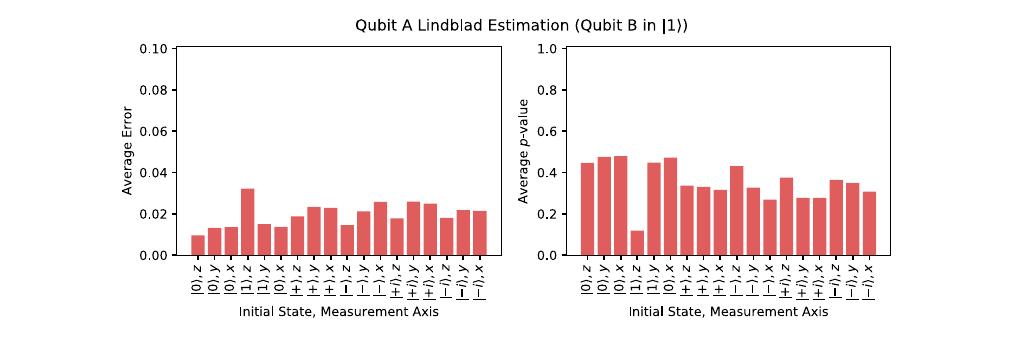}
	\includegraphics[scale=1]{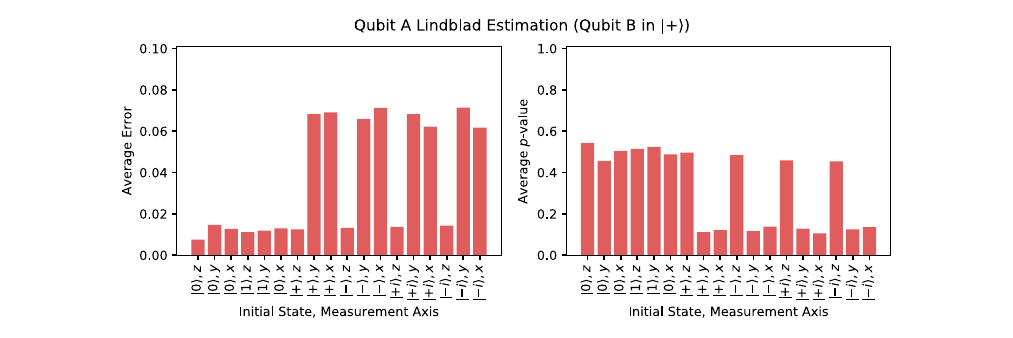}
	\caption{Analyzing the Lindblad extraction of qubit A's idling channel when qubit B is in $\ket{0}$, $\ket{1}$, and $\ket{+}$. The error and $p$-value between the data and the Lindblad estimation are calculated for each time step, initial state, and measurement axis of qubit A, and the results are averaged over the first 20$\mu s$. Note that, when qubit B is prepared in $\ket{+}$ (bottom plots), we observe dramatically larger errors and lower $p$-values for the sequences corresponding to $T_2$-like measurements of qubit A, consistent with the poor Lindblad fits shown in Fig.~\ref{fig:1QB_full_LT}c of the main text. Meanwhile, the sequences corresponding to $T_1$-like measurements of qubit A (which are blind to the entanglement with qubit B) show comparatively good fits to data even when qubit B is in $\ket{+}$.}
	\label{fig:figurePREsingleL}
\end{figure*}

\begin{figure*}[t!]
    \centering
    \includegraphics[scale=1]{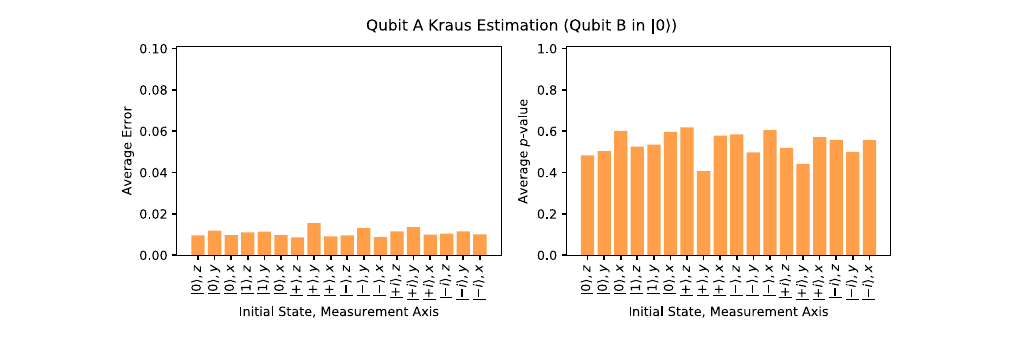}
    \includegraphics[scale=1]{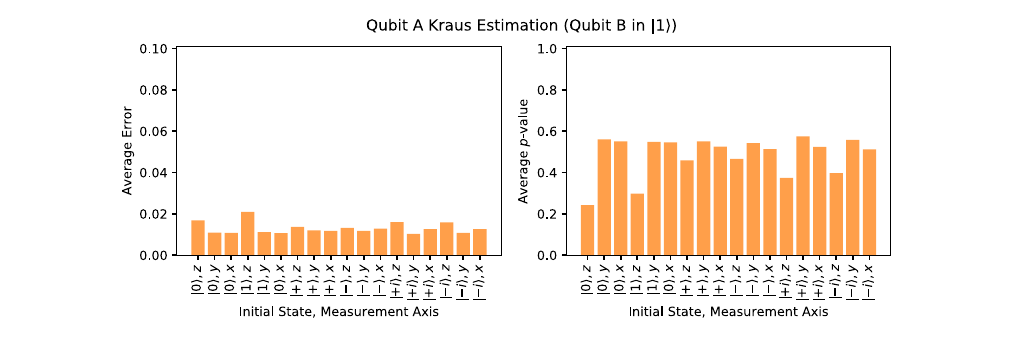}
    \includegraphics[scale=1]{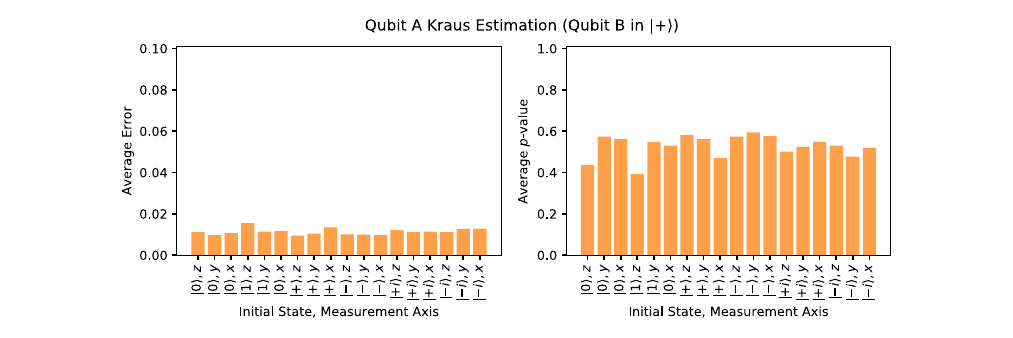}
    \caption{Analyzing the Kraus extraction of qubit A's idling channel when qubit B is in $\ket{0}$, $\ket{1}$, and $\ket{+}$. The error and $p$-value between the data and the Kraus estimation are calculated for each time step, initial state, and measurement axis of qubit A, and the results are averaged over the first 20$\mu s$.}
    \label{fig:figurePREsingleK}
\end{figure*}

\begin{figure*}[t!]
	\centering
    \includegraphics[scale=1]{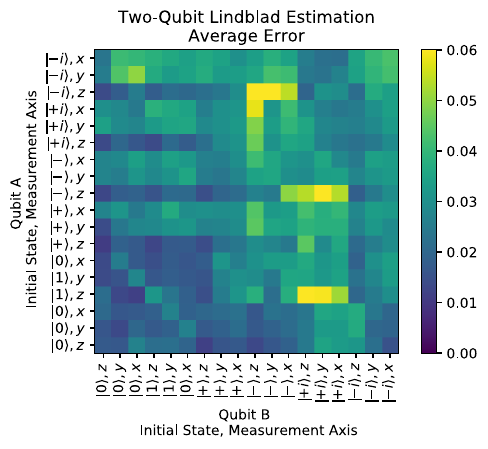}
	\includegraphics[scale=1]{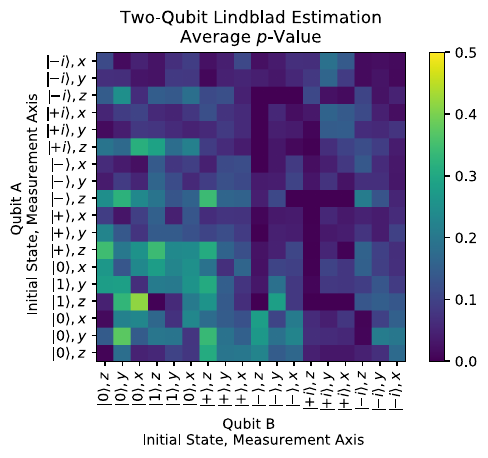}
	\caption{Analyzing the Lindblad extraction of the two-qubit idling channel. The error and $p$-value between the data and the Lindblad estimation are calculated for each time step and qubit configuration (qubit A on y-axis, B on x-axis), and the results are averaged over the first 20$\mu s$.  
    %Note the outlier in the bottom left corner of the RE grid: for this measurement, the expected probability of measuring $\ket{11}$ is close to 0---making calculation of the relative error extremely sensitive---and we compute a relative error of  $\sim$56\%.  The remaining grid values are less than 28\%.
    }
	\label{fig:figurePREtwolind}
\end{figure*}

\begin{figure*}[t!]
	\centering
    \includegraphics[scale=1]{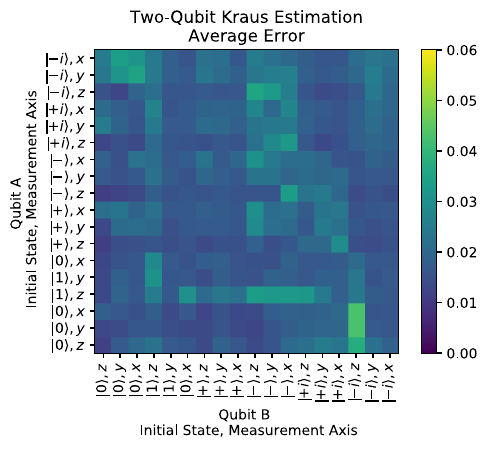}
	\includegraphics[scale=1]{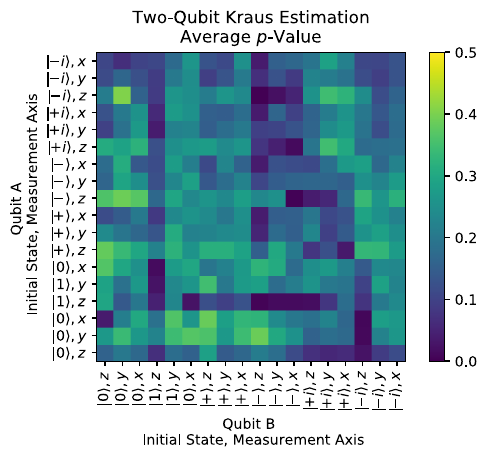}
	\caption{Analyzing the Kraus extraction of the two-qubit idling channel. The error and $p$-value between the data and the Kraus estimation are calculated for each time step and qubit configuration (qubit A on y-axis, B on x-axis), and the results are averaged over the first 20$\mu s$.  }
	\label{fig:figurePREtwokraus}
\end{figure*}

\begin{figure*}[t!]
	\centering
	\includegraphics[scale=1]{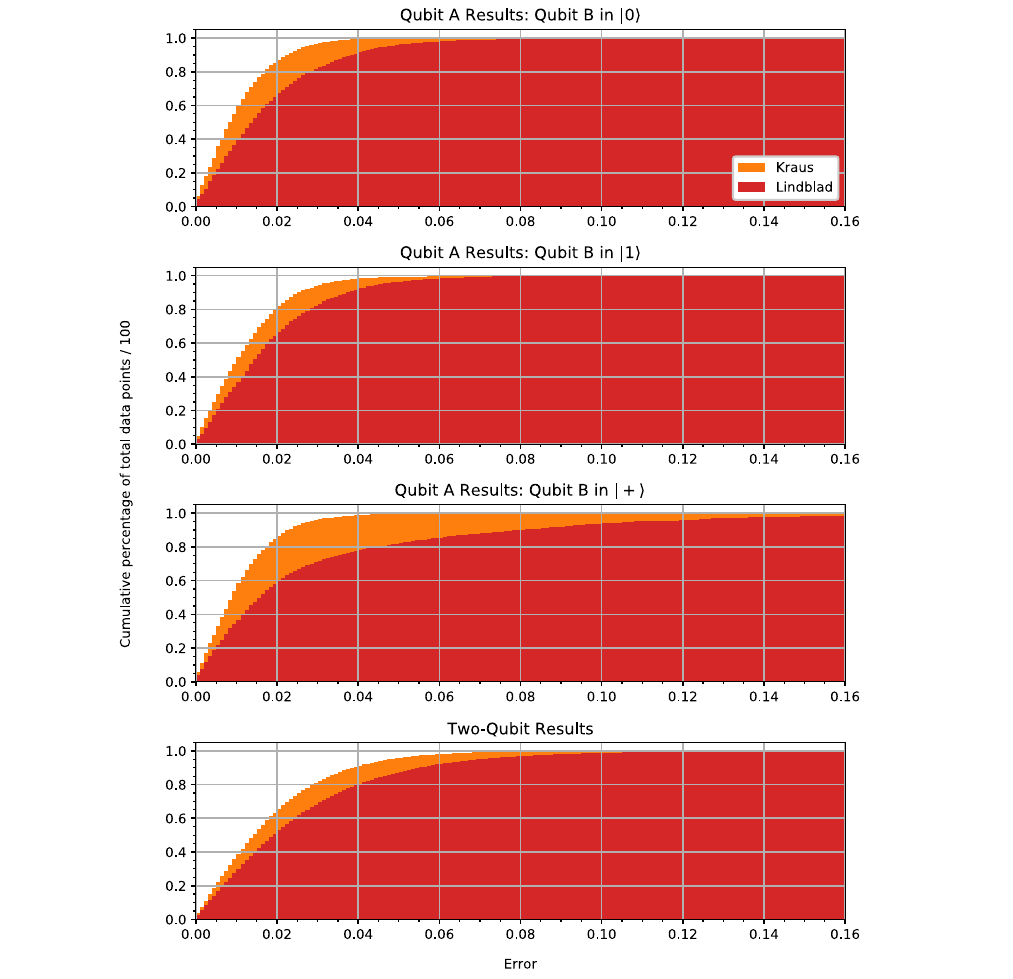}
	\caption{Cumulative histogram of the error for each data point.  Instead of averaging over all time steps, the cumulative histograms treat each combination of time step, initial state, and measurement axis as a single data point and bins the error between model (Kraus in orange, Lindblad in red) and data at each point. The data plotted here are compiled from all time steps (i.e., not just the first 20$\mu$s) for each of the four cases in Fig.~\ref{fig:figure5} of the main text.  For example, looking at the results of all two-qubit data estimated with the Lindblad MLE (bottom plot, red), we find that 80\% of the predicted probabilities fall within 0.04 of the measured probabilities.  The Kraus estimates yield better fits most of the time (since, unlike the Lindblad estimates, they are not constrained to any Markovian model), and thus they converge to the 100\% limit more rapidly.}
	\label{fig:figurePercentiles}
\end{figure*}

In this section, we provide additional statistical analysis of the single- and two-qubit operators reported in the main text.  For a single qubit, we can model the results for each input state, evolution time, and measurement basis as a Bernoulli distribution (a biased coin toss yielding either `0' or `1'), from which we obtain 1000 independent shots.  The same can be said of the two-qubit data, with the results corresponding to a multinomial distribution with four outcomes (`00', `01', `10', and `11').  The goal of LT is to find a set of model parameters that closely estimates these outcome probabilities across the entire set of initial states, measurement axes, and time steps.  To evaluate our success in finding such a model, we compute two standard metrics to quantify the deviation between the model predictions and data: the absolute error and the chi-square goodness of fit ($\chi^2$).  

The error between the model and data at each time step is computed as 
\begin{align}\label{eq:RE}
	\text{error}(t_i) = \Big| x^{\text{meas}}_i - x^{\text{model}}_i \Big| %\times 10^2
\end{align}
where $x_i^{\text{meas}}$ is the measurement probability obtained in experiment at time step $t_i$ (blue dots in Fig.~\ref{fig:1QB_full_LT},\ref{fig:figure4} of the main text) and $x_i^{\text{model}}$ is the corresponding estimate from the outcome of the MLE routine (orange $\times$'s in Fig.~\ref{fig:1QB_full_LT},\ref{fig:figure4} for the Kraus extraction, red line for the Lindblad).  

The $\chi^2$ test is a statistical quantifier of how likely it is that the data could have been produced by an assumed model known as the null hypothesis.  Our null hypothesis shall be the assumption that the data is well fit by a time-independent Markovian master equation.  The $\chi^2$ value is then computed as 
\begin{align}
	\chi^2(t_i) = \sum_{j=1}^{k} \frac{(x_{i,j}^{\text{meas}} - x_{i,j}^{\text{model}})^2}{x_{i,j}^{\text{model}}}
\end{align}
with the same definitions of $x^{\text{meas}}$ and $x^{\text{model}}$ as in Eq.~\eqref{eq:RE}.  However, unlike in the error calculation, we must also sum over the the total number of categories $k$ that the data can fall into: 2 for the single-qubit data (`0' or `1'), 4 for the two-qubit data (`00', `01', `10', or `11').  Under the null hypothesis, the deviation between $x_i^{\text{observed}}$ and $x_i^{\text{expected}}$ is normally distributed due to the central limit theorem, and it is well-known that (given a large enough sample size\footnote{It is acceptable to use the chi-square test instead of an exact test, such as the binomial or multinomial test, when the number of counts in any observed category (i.e., measurement outcome) is not too low, around 5 counts~\cite{Zar2007}.  We find that the lowest number of counts of some type across all single-qubit data is 146, and across all two-qubit data is 9.}) the $\chi^2$ statistic follows the $\chi^2$-distribution with $k-1$ degrees of freedom~\cite{Cochran1952}.  Intuitively, the larger the value of $\chi^2$, the greater the discrepancy between the observed and expected values.  For any value of $\chi^2$, the $\chi^2$-distribution can then be used to compute the probability that a value at least as extreme might have been obtained, which is known as the $p$-value.  If one finds a particularly small $p$-value, then one should consider rejecting the null hypothesis on the grounds that the assumed model is not very likely to have actually produced the observed data.\footnote{It is important to stress that one should \textit{not} assume the converse, that the null hypothesis is wrong with a probability given by the $p$-value.  The $p$-value also says nothing about what the correct hypothesis might actually be in the case that the null hypothesis is rejected, just that a more plausible alternative likely exists.}  Whether one accepts or rejects the null hypothesis is determined by a threshold $p$-value (commonly denoted as $\alpha$) which is chosen in advance of analyzing the data.  This threshold is set arbitrarily and simply expresses how conservative one would like to be when deciding to reject the null hypothesis, trading off false positives for false negatives as the threshold is set lower and lower.  In our case, we will refrain from choosing a specific $\alpha$ and let the data speak for itself,  noting that higher $p$-values indicate data that is more compatible with a Markovian assumption, while lower values suggest deviation between the Markovian model and data.

In the following analysis, we compute the average error and $p$-value over time, for each combination of initial state and measurement basis.  The $p$-value is computed from the $\chi^2$ distribution with one degree of freedom to account for the two different measurement outcomes (three degrees of freedom for the two-qubit data, $k-1$ in general). Since the different outcomes are not inherently included in the calculation of the error, we also average over the different outcome types when calculating error.  Since we are calculating the error between two probabilities, the error is bounded between 0 and 1. The $p$-value is inherently bounded between 0 (bad fit) and 1 (exact fit).  Under the null hypothesis, the $p$-values should actually be uniformly distributed due to statistical error, and so an average $p$-value of around 0.5 indicates very close agreement with the null-hypothesis.  We emphasize that in all cases, whenever the expected probabilities are small, the relative error and the $p$-value will both suffer even if the fit is qualitatively quite good.  All statistics are computed based on data drawn within the first 20$\mu$s, corresponding to the most coherent part of the evolution, where non-Markovian effects are most apparent.

In Figs.~\ref{fig:figurePREsingleL} and \ref{fig:figurePREsingleK}, we plot the average error and $p$-values for the Lindblad and Kraus extraction of qubit A's idling channel when qubit B is in $\ket{0}$, $\ket{1}$, and $\ket{+}$. When qubit B is prepared in $\ket{0}$ or $\ket{1}$ (i.e., the cases where we would expect the Markovian assumption to hold for qubit A's channel), the $p$-values are on average above 0.2, and the error values are always less than 0.03, except for two outlier cases: prepare $\ket{+}$, measure in $y$, qubit B in $\ket{0}$ (Lindblad fit shown in Fig.~\ref{fig:figurelindblad1}); prepare $\ket{1}$, measure in $z$, qubit B in $\ket{1}$.  In comparison, when qubit B is prepared in $\ket{+}$ and we perform a $T_2$-like measurement on qubit A (initialize and measure in the $x$- or $y$-bases), we find a lower $p$-value (less than 0.15) and a large error (greater than 0.06).  As shown in the main text, these cases correspond to evolution whose actual precession frequencies are not well-predicted by the Lindblad fit, due to entanglement with qubit B.  It is worth noting that, even in this non-Markovian scenario, there are still sequences that are well-fit by a Lindbladian: these sequences (which correspond to $T_1$-type measurements) are blind to the entanglement produced by the $ZZ$-coupling between the qubits, and it follows that they can be fit to a Markovian model.  
% For completeness, in Fig.~\ref{fig:figurePREexcluding} we show average $p$-values and RE values for the two exclusion cases shown in Figures \ref{fig:figurelindblad1_excl1} and \ref{fig:figurelindblad1_excl2}.  
In Fig.~\ref{fig:figurePREtwolind} and ~\ref{fig:figurePREtwokraus}, we provide the same analysis for the two-qubit Lindblad and Kraus estimates respectively, where the initial state and measurement basis for qubits A and B are labeled along the axes.  Lastly, in Fig.~\ref{fig:figurePercentiles}, we plot cumulative histograms of the error values for the four datasets emphasized in Fig.~\ref{fig:figure5} of the main text, providing a high-level view of the fit-quality across all data points.

%

% \bibliography{Lindblad_Tomography}

\end{document}